\documentclass[a4paper, 10pt, leqno]{amsart}

\usepackage{amsmath,amsfonts}
\usepackage{amsmath}
\usepackage{graphicx}
\usepackage{rotating}
\usepackage{caption}
\usepackage{fancyhdr}
\usepackage{hyperref}
\usepackage{amsaddr}
\usepackage{subcaption}

\begin{document}
\pagestyle{plain}
\bibliographystyle{plain}
\newtheorem{theo}{Theorem}[section]
\newtheorem{lemme}[theo]{Lemma}
\newtheorem{cor}[theo]{Corollary}
\newtheorem{defi}[theo]{Definition}
\newtheorem{prop}[theo]{Proposition}
\newtheorem{problem}[theo]{Problem}
\newtheorem{remarque}[theo]{Remark}
\newcommand{\beq}{\begin{eqnarray}}
\newcommand{\enq}{\end{eqnarray}}
\newcommand{\be}{\begin{eqnarray*}}
\newcommand{\en}{\end{eqnarray*}}
\newcommand{\ben}{\begin{eqnarray*}}
\newcommand{\enn}{\end{eqnarray*}}
\newcommand{\Td}{\mathbb T^d}
\newcommand{\Rd}{\mathbb R^n}
\newcommand{\R}{\mathbb R}
\newcommand{\N}{\mathbb N}
\newcommand{\Sn}{\mathbb S}
\newcommand{\Zd}{\mathbb Z^d}
\newcommand{\Linf}{L^{\infty}}
\newcommand{\dt}{\partial_t}
\newcommand{\Dt}{\frac{d}{dt}}
\newcommand{\Dtt}{\frac{d^2}{dt^2}}
\newcommand{\demi}{\frac{1}{2}}
\newcommand{\vf}{\varphi}
\newcommand{\epu}{_{\varepsilon}}
\newcommand{\ep}{^{\varepsilon}}
\newcommand{\bfi}{{\mathbf \Phi}}
\newcommand{\bpsi}{{\mathbf \Psi}}
\newcommand{\bx}{{\mathbf x}}
\newcommand{\dis}{\displaystyle}
\newcommand{\ds}{\partial_s}
\newcommand{\dss}{\partial_{ss}}
\newcommand{\dx}{\partial_x}
\newcommand{\dxx}{\partial_{xx}}
\newcommand{\dy}{\partial_y}
\newcommand{\dyy}{\partial_{yy}}
\newcommand {\g}{\`}
\newcommand{\E}{\mathbb E}
\newcommand{\bQ}{\mathbb Q}
\newcommand{\1}{\mathbb I}
\newcommand{\bF}{\mathbb F}
\newcommand{\F}{\cal F}
\newcommand{\bP}{\mathbb P}
\let\cal=\mathcal
\newcommand{\lb}{\langle}
\newcommand{\rb}{\rangle}
\newcommand{\bv}{\bar{v}}
\newcommand{\uv}{\underline{v}}
\newcommand{\cS}{{\cal S}}
\newcommand{\cSf}{\cS_\Phi}
\newcommand{\bw}{\bar{w}}
\newcommand{\uw}{\underline{w}}
\newcommand{\bsig}{\bar{\sigma}}
\newcommand{\usig}{\underline{\sigma}}


\author{Gr\'egoire LOEPER} 
\address{Monash University, School of Mathematical Sciences}
\email{gregoire.loeper@monash.edu}
\title{Option pricing with linear market impact and non-linear Black-Scholes Equations}
\date{\today}
 

\begin{abstract}
We consider a  model of linear market impact, and address the problem of replicating a contingent claim in this framework.
We derive a non-linear  Black-Scholes Equation that provides an exact replication strategy.

This equation is fully non-linear and singular, but we show that it is well posed, and we prove existence of smooth solutions for a large class of final payoffs, both for constant and local volatility. To obtain regularity of the solutions, we develop an original method based on Legendre transforms.

The close connections with the problem of hedging with {\it gamma constraints} \cite{SonTouz}, \cite{SonTouz2}, \cite{CheriSonTouz},  with the problem of hedging under {\it liquidity costs} \cite{CetinSonerTouzi} are discussed. The optimal strategy and associated diffusion are related with the  {\it second order target problems} of \cite{SonTouzZhang}, and with the solutions of {\it optimal transport problems by diffusions} of \cite{TanTouz}.

We also derive a modified Black-Scholes formula valid for asymptotically small impact parameter, and finally provide numerical simulations as an illustration.
\end{abstract}

\maketitle

\section{Introduction}
This paper is about the derivation and mathematical analysis of a pricing model that takes into account the market impact of the option's hedger, i.e. the feedback mechanism between the option's delta-hedging and the price dynamics.
We will throughout the paper assume a linear market impact: 
each order to buy $N$ stocks impacts the stock price by $\lambda N S^2$ ($\lambda \geq 0$). This scaling, that will be discussed hereafter, means that the impact in terms of relative price move depends on the amount of stock traded expressed in currency, hence $\lambda$ is homogeneous to the inverse of the currency {\it i.e. percents per dollar}. 
\beq\label{fondam}
\hbox{Order to buy N stocks} \implies  S \to S(1+\lambda NS).
\enq
We will examine several situations where $\lambda$ is either constant, or can be also a function of the solution itself.

The literature  devoted to the study of the market impact itself is quite vast, and  it is more frequent to find an impact that varies as a power law of the size of the trade (see for example \cite{Almgren}).
Although our linear approach is clearly not the most realistic in terms of market microstructure, it has the advantage of avoiding arbitrage opportunities, as well as not being sensitive to the hedging frequency: in a non-linear model, splitting an order in half and repeating it twice would not yield the same result, a situation that we want to avoid here, as we aim at deriving a time continuous formulation.

In addition we assume that the market impact is permanent: there is no relaxation following the immediate impact, where the price goes back partially to its pre-trade value. The case with relaxation has been studied in a companion paper \cite{AbergelLoeper}.

\subsection{Order book modeling}
From an order book perspective, our market impact model consists in assuming an order book with continuous positive density around the mid-price, and where following a market-order starting at price $S$ (say a buy order for example), where liquidity has been consumed up to $S+\delta S$, the new mid-price becomes $S+\delta S$, and the liquidity removed between $S$ and $S+\delta S$ is instantaneously refilled with buy orders (or sell orders if the price had moved down after a sell order). Clearly this model is a great simplification of what is actually observed in real markets (see \cite{BouchaudMI} for instance), but as we will show this simple approach is enough to obtain a non-trivial modification of the usual Black-Scholes equation. More precisely:
Assume a static order book, parametrized by a mid price $\bar S$ and a {\it supply} intensity $\mu(t,s)$ in the following way: the number of stocks available for purchase between $S$ and $S+dS$ equals to $\mu(t,S)dS$. 

We assume that {\it the (local) liquidity (expressed in currency) available between $\bar S$ and $\bar S(1+\varepsilon)$ equals $L(t,\bar S) \varepsilon + o(\varepsilon)$.}
Simple calculations show that the scaling (\ref{fondam}) implies that $L(t,S)$ is actually constant, which explains our choice, as we believe that the quantity $L$ is the good measure of a stock's liquidity. Then independently of this choice, when spending an amount $A$ (expressed in currency) in stocks, the order book will be consumed up to $\bar S(1+\epsilon)$ given by
$$A = \int_{\bar S}^{\bar S(1+\varepsilon)} \mu(t,s) s ds,$$
while the number of stocks purchased will be
$$N = \int_{\bar S}^{\bar S(1+\varepsilon)} \mu(t,s)ds.$$
One sees right away that, at the leading order, the average price of execution is $\bar S (1+\demi\varepsilon)$. 
In a subsequent study \cite{AbergelLoeper}, the authors consider an immediate {\it  relaxation} of the price after the liquidity is consumed, hence the liquidity of the order book is rebuilt around $\bar S(1+\gamma \varepsilon)$ for a certain relaxation factor $\gamma \in [0,1]$. We will only adress here the case $\gamma=1$. The choice $\gamma=0$ is the one that has been studied by Cetin, Soner and Touzi in \cite{CetinSonerTouzi}: no permanent market impact, but liquidity costs.

\subsection{Motivations and links with previous works}
In terms of concrete applications, the problem of derivatives pricing with market impact arises when the delta hedging of the option implies a volume of transactions on the underlying asset  that is non-negligible compared to the average daily volume traded.
For example, a well observed effect known as {\it stock pinning} arises when the hedger is long of the (convex) option for a large notional, and one observes then a decrease of realized volatility if the underlying ends near the strike at maturity.
In financial terms, the hedger of the option makes a loss if the volatility realizes below its implied value. This stylized fact can be recovered by our pricing equation. Conversely, when selling a convex option for a large notional, a common market practice on derivatives desks is to super-replicate the option by the cheapest payoff satisfying a constraint of gamma (the second derivative of the option), the {\it gamma max} being adjusted to the liquidity available on the option's underlying.
Hence there are two issues arising here: 
\begin{itemize}
\item[-] Be able to control a priori the trading volume due to delta-hedging: this approach has been studied in depth in the works
 by Soner and Touzi \cite{SonTouz2} and Cheridito, Soner and Touzi \cite{CheriSonTouz} that deal with the problem of heging with gamma-constraints. 
\item[-] Quantify the liquidity costs induced by the delta-hedging, and incorporate them in the option's price, this has been studied by Cetin, Jarrow and Protter in \cite{CetinJarrowProtter}  and by Cetin, Soner and Touzi in \cite{CetinSonerTouzi}. However these works consider only liquidity costs, and not the effect of permanent market impact.
\end{itemize}
Our approach addresses those two issues: first, via the market impact mechanism, it induces a constraint on the gamma when selling a convex payoff,  and it constraints the theta (the time derivative of the option)  when selling a concave payoff.
Thus it recovers two important stylized facts of the gamma constraint approach, moreover it incorporates liquidity costs in the price. It can be noted that the parabolic operator that we obtain through market impact lies somehow between the gamma-constraint operator of \cite{SonTouz} and the liquidity costs operator of \cite{CetinSonerTouzi}.

A formal argument  shows that, playing on the dependency of the market impact parameter with the solution, one can recover exactly the gamma constraint pricing equation of Cheridito Soner Touzi \cite{CheriSonTouz}, or the liquidity cost equation of Cetin, Soner and Touzi in \cite{CetinSonerTouzi}. This can be seen as closely related to  the work of Serfaty and Kohn  \cite{SerfatyKohn} that recover non-linear heat equations by stationary games approach. From a financial modelling perspective, this would amount to assume a supply curve for the price of gamma-hedging.

 Concerning the mathematical techniques, as opposed to \cite{CheriSonTouz}, \cite{CetinSonerTouzi}, \cite{SonTouz2}, \cite{BoLoZo1}, \cite{BoLoZo2}, we adress the problem of finding exact replication strategies, while the aforementioned work deal with stochastic target problems: {\it find the cheapest trading strategy that super-replicates the final payoff}. The solutions of these two problems co\"incide in general, but may differ in some degenerate cases (see \cite{SonTouz3}, \cite{AbergelLoeper}). 
Apart from introducing the linear market impact model (which has then led to the subsequent studies \cite{BoLoZo1}, \cite{BoLoZo2}, \cite{AbergelLoeper}), the main contribution of this paper is a complete study of a fully non linear parabolic equation associated to a new class of stochastic control problems  (see Theorem \ref{theocstvol}).  We will also prove a representation formula that gives some qualitative informations about the modified dynamics. Those two results are then used to derive rigorously an asymptotic expansion of the solution for small market impact, leading to a modified Black-Scholes-{\it Legendre} formula, as well as a simple and efficient way of computing the market impact effect.

We believe that the techniques used toward obtaining these results are original and of independent interest. As will be discussed hereafter, these results can be seen as related to regularity results concerning porous media equations (see  the book of Vazquez \cite{VazquezDecay} for a complete reference), in particular in the case of {\it Fast Diffusion Equations}, as well as
the papers by Crandall and Pierre \cite{CranPieDelta}, \cite{CranPieA} about interior regularity for  non-linear diffusions.

 The stochastic target problem associated to our model has been studied in two companion papers \cite{BoLoZo1}, \cite{BoLoZo2}, where it is shown that the exact replication strategy is actually the optimal solution for the stochastic target problem. To make a connection between the two results, it is important to notice that through stochastic control techniques, one is able to derive a viscosity formulation of the value function. It is only when this function has enough regularity that one can deduce the optimal strategy from the value function, hence the importance of the question of regularity.



Finally we mention the references (\cite{lambertonPhamSchweizer}, \cite{platenSchweizer}, \cite{schonbucherWilmott}, \cite{frey}, \cite{freyStremme}, \cite{platenSchweizer}) that  also address the problem of option hedging in non-perfect markets with different approaches, leading to different mathematical techniques.

\subsection{The pricing equation}
As we will see,  assuming no interest rates and dividends, the pricing equation that we obtain can be put under the form
\beq \label{newbs}
\dt u + \demi\sigma^2\frac{s^2\partial_{ss}u}{1-\lambda s^2\partial_{ss}u}=0,
\enq
which reads also 
\ben 
\frac{\sigma^2}{2\partial_t u} + \frac{1}{s^2\partial_{ss}u} = \lambda,
\enn 
where $\lambda$ can be either constant or dependent of the solution as $\lambda(s^2\dss u)$. In this case we obtain a wide class of fully non-linear Black Scholes equations of the form
\be 
\dt u + \demi \sigma^2 F(s^2\partial_{ss}u)=0.
\en
As will be shown, one can derive any  parabolic equation of this form through an ad-hoc choice of $\lambda$, as long as $F(\gamma)\geq \gamma$ and $F(0)=0$.

The case of equation (\ref{newbs}) is quite challenging for the mathematical perspective: the operator $\dss u \to F(s^2\dss u)$  is not uniformly elliptic (when $\dss u$ goes to $-\infty$), and is singular (when $s^2\dss u$ goes to $\lambda^{-1}$). Standard theory does not apply right away, and an ad-hoc regularity theory must be developed. Still, we will be able to show interior (i.e. not relying on a smooth terminal payoff) regularity of solutions for constant $\lambda$ and boundary regularity (i.e. assuming a good terminal condition) for non-constant $\lambda$.

Let us rewrite equation (\ref{newbs}) as follows:
\ben
&&\dt u + \demi\sigma^2F(s^2\dss u)=0,\\
&&F(\gamma)= \frac{\gamma}{1-\lambda\gamma}=\frac{1}{\lambda}(-1+\frac{1}{1-\lambda\gamma}).
\enn
The link with non linear diffusions appears when one differentiates twice the equation: assuming $\sigma$ constant one then obtains for $\beta=1-\lambda s^2\dss u$
\ben
\dt \beta - \frac{\sigma^2  s^2}{2}\dss\left(\frac{1}{\beta}\right)=0,
\enn
which can be seen  as a lognormal version of the {\it Fast Diffusion Equation} (see \cite{VazquezDecay}) 
\ben
\dt u - \Delta\left(\frac{1}{u}\right)=0.
\enn


\subsection{Second order target problems and optimal transport by diffusions}
There is also a clear connection between this work and the work by Soner and Touzi and Zhang \cite{SonTouzZhang} about dual formulation of second order target problems, and also with the problems of optimal  transport
 by controlled martingales studied in \cite{TanTouz}. Indeed observe that the elliptic operator $F(\gamma)$ in (\ref{newbs}) is convex, and then applying the reasoning of \cite{SonTouzZhang} based on the Legendre transform representation of $F$, one can show formally that a solution to (\ref{newbs}) will also be solution of the following variational problem:
\beq\label{variat}
u(t,s)=\sup_{a \in {\cal A_T}}\Big\{ \E\{\Phi(S_T) - \frac{1}{2\lambda}\int_t^T (a_{t'}^{1/2}-\sigma(t',S^a_{t'}))^2dt'\Big\}.
\enq
where $S^a$ starts from $s$ at time $t$ and follows $dS^a_r= a^{1/2}_r S^a_r dW_r$, $W_r$ is a Brownian motion,
 ${\cal A_T}$ is the set of all  bounded positive adapted processes on $[t,T]$, and $\sigma(t,s)$ is given and positive. 
 
 This can be seen as a stochastic version of Hopf-Lax formula, 
\beq\label{hopf-lax}
u(t,s)=\sup_{\gamma \in {C^1([0,T])}, \gamma(t)=s}\Big\{ \Phi(\gamma(T)) - \frac{1}{2\lambda}\int_t^T H(\dt\gamma(t'))dt'\Big\},
\enq 
see \cite{Ba}, which is a representation formula for solutions to 
\ben
\dt u + H^*(D_x u)=0,
\enn
where $H^*$ is the Legendre transform of $H$. 
The fact that, in general, the solution of a variational problem like (\ref{variat}) is indeed the viscosity solution of an associated parabolic equation (which here would be \ref{newbs}) has been studied in \cite{SonTouzZhang}. In this particular case we will be able to actually characterize the optimal $\hat{a}$ in (\ref{variat}), and  show that the optimal diffusion $S^{\hat{a}}$ is a martingale up to time $T$ under mild conditions, which allow $\hat{a}$ to be unbounded.

Then, still reasoning formally,  $S^{\hat{a}}$ transports its initial distribution on its final distribution minimizing the transport cost
\ben
\E\left(\frac{1}{2}\int_0^T (a_t^{1/2}-\sigma(t,S_{t}))^2dt\right),
\enn
which is a particular case of the problem studied in \cite{TanTouz}. More generally the link between the variational problem (\ref{variat}) and optimal transport arises through the use of Kantorovitch duality, see \cite{kanto1}, \cite{kanto2}, \cite{Br1} and also \cite{Vi} for an overview on optimal transport, and also \cite{L3} for further extensions to non-linear variational problems.

 The regularity results that we obtain here will provide the regularity of the optimizers in the aforementioned  problems.

\subsection{Organization of the paper}
The rest of the paper is organized as follows: in the next section (Section 2) we provide a heuristic derivation of the pricing equation, and discuss its practical  relevance for financial modelling.

Section 3 gives the time continuous formulation of the problem as a system of stochastic differential equations, and shows that the pricing equation (\ref{newbs}) actually leads to an exact replication strategy when it admits a smooth solution, see Theorem \ref{theo_verif}. 

Section 4 establishes the connection between the solution of (\ref{newbs}) and the problem (\ref{variat}), and discusses also the link with optimal transport and robust hedging.

Section 5 contains the  regularity results for the pde (\ref{newbs}). Theorems \ref{theocstvol}, \ref{ubvbm} and \ref{theolocvol},  contain the main a priori estimates for solutions to (\ref{newbs}), that hold under mild conditions on the final payoff. 
These results lead then to the existence, regularity and uniqueness of the solutions to (\ref{newbs}): Theorem \ref{theo-main}.

Section 6 establishes representation formulas for the solution: Theorems \ref{local-rep}, \ref{local-rep-2} and \ref{maintheo}. These representation formulas are a probabilistic counterpart to the pde (\ref{newbs}), and give  qualitative insight about the solution, and establish further results concerning the optimizers of the problem (\ref{variat}). 

In section 7 we derive a first order expansion of the solution for small market impact ($\lambda$). This leads also to modified Black-Scholes-{\it Legendre} formula (see formulas (\ref{newBS}, \ref{BSformula}, \ref{BSformulaput})), which is easily computed by analytic formulas or standard Monte-Carlo simulations.
It should formally be also valid in the liquidity costs case of \cite{CetinSonerTouzi}.

Section 8 illustrates the paper with a numerical solution of the pde.

\subsection{Acknowledgements}
This paper has been in gestation for a long time, its motivation originally appeared while the author was a member of the Quantitative Research team at BNP Paribas Capital Markets, and I thank my colleagues there for stimulating discussions. Part of this work has also been done under the hospitality of the Chair of Quantitative Finance at the Ecole Centrale de Paris, and I thank in particular Fr\'ed\'eric Abergel. I also thank Bruno Bouchard, Nizar Touzi, Juan Vazquez, Fernando Quiros, and Fima Klebaner for enlightening discussions.

\section{Heuristics}
We assume that we have {\bf sold} an option whose value is $u(t,s)$, and greeks are as usual 
\be
\Delta&=&\partial_s u,\\
\Gamma&=&\partial_s\Delta,\\
\Theta&=&\partial_t u.
\en
We also introduce the Gamma {\it in currency}, i.e. 
\beq\label{gammac}
\gamma = \Gamma S^2.
\enq

The strategy that we use is the following: we assume a priori that there exists an exact replication strategy, that consists in holding $\Delta=\partial_s u$ stocks. We compute the equation that must be followed by $u$, and the modified dynamics that this strategy implies. We then check that this strategy allows indeed to perfectly replicate the final claim.

Starting from a delta-hedged portfolio, assume that the stock price $S$ moves by $dS$. 
We will assume that this initial move of $dS$ is given as is usual by
\ben
dS=S\sigma dW_t,
\enn
where $W_t$ is a standard Brownian motion, and $dW_t$ its increment between $t$ and $t+dt$ (those objects will be introduced more formally later on). 
A `naive' hedge would be to buy $\Gamma dS$ stocks, but as this order will impact the market, the portfolio will not end up delta-hedged. 
Assume instead a hedge adjustment of $\mu \Gamma dS$ stocks, and let us  find $\mu$ such that spot ends  up at a final value $S+\mu dS$. Using (\ref{fondam}), we write that

\be
S_{\text{after re-hedging}} - S_{\text{before re-hedging}} = \lambda S^2 \;\text{Number of stocks bought to re-hedge},
\en
and this yields
\be
\mu dS - dS = \lambda [\Delta(S+\mu dS)-\Delta(S)] S^2 .
\en
This identity expresses the fact that the number of titles  bought is $\Delta(S+\mu dS)-\Delta(S)$, hence that the portfolio is delta-hedged at the end of the trade.
Performing a first order Taylor expansion leads to
\beq\label{taylor}
\Gamma \mu dS \lambda S^2 = (\mu- 1) dS + o(dS),
\enq
which yields
\beq\label{defmu}
\mu = \frac{1}{1-\lambda \gamma}.
\enq

Remember that $\gamma$ is computed with respect to the option the portfolio is short of, hence $\gamma > 0$ when one sells a call for example. Assuming that $\lambda\gamma < 1$  (to be discussed later), one sees that $\mu>1$: as expected the hedger increases the volatility by buying when the spot rises, and selling when it goes down.

One can also reach the conclusion (\ref{defmu}) by following an iterative hedging strategy: after the initial move $S \to S+dS$, the hedge is adjusted "naively" by $\Gamma dS$ stocks, which then impacts the price by $dS_2 = \lambda S^2\Gamma dS$. A second re-hedge of $\Gamma dS_2$is done,  which in turn impacts the price and so forth. The final spot move is thus the sum of the geometric sequence
\ben
dS(1 + \lambda S^2\Gamma + (\lambda S^2\Gamma)^2 + ...) = \frac{dS}{1-\lambda\gamma}.
\enn
One sees right away that in this sequence, the critical point is reached when the "first" re-hedge (i.e. buying $\Gamma dS$ stocks after the initial move of $dS$) doubles the initial move: the sum will not converge. This situation will be discussed hereafter.

Then the value $Q$ of the portfolio containing -1 option + $\Delta$ stocks at the beginning of time $t$, and $\Delta(S+\mu dS)$ stocks at time $t+dt$ evolves as
\be
dQ = u(t,S) - u(t+dt, S+\mu dS) + \Delta \mu dS + \cal{R}
\en
and $\cal{R}$  is the profit realized during the re-hedging. In a perfect frictionless market, this term is zero, as one buys $d\Delta$ stocks at a price $S+dS$, and the "post-re-hedge" value of the stocks is $S+dS$.
To compute ${\cal R}$  we recall that
\beq
\label{fondambis1}\hbox{Immediate impact of an order to buy N stocks} &\implies & S \to S(1+\lambda NS),  \\
\label{fondambis2} \text{Average execution price }& = & S(1+\demi\lambda NS).\enq
The computation of $\cal{R}$ then yields
\ben
\cal{R} &=& N\Big(\text{Final price of the stocks bought} - \text{Average execution price}\Big) \\
&=& N\Big(S(1+\lambda NS) - S(1+\demi \lambda NS)\Big)\\
&=&\demi\lambda N^2S^2,
\enn
with $N=\Gamma\mu dS$.
Finally if $V$ is the the value of the hedger's trading strategy, and setting $d\tilde S = \mu dS$, one obtains:
\beq\label{dVheuristics}
dV = \Delta d\tilde S + \demi \lambda S^2(\Gamma d\tilde S)^2.
\enq

\subsubsection{The pricing equation}

We now assume that the option is sold at its fair price, hence $dQ = 0$ at first order in time. 
Then we have as $S$ moves to $S+\mu dS$,
\be
du = \dt u  dt + \partial_s u \mu dS + \demi \partial_{ss}u (\mu dS)^2 + o(dt),
\en
and we thus get 
\be
\dt u dt + (\mu dS)^2/2\left[\partial_{ss}u - \lambda(\partial_{ss}u)^2S^2 \right]=o(dt),
\en
with $\mu$ defined by (\ref{defmu}). We recall now that the {\it initial} move of $S$ i.e. $dS$ in our notations, is driven by a geometric Brownian motion, as in the Black-Scholes model, so $dS=S\sigma dW$. Then one can following It\^o's formula replace $(dS)^2$ by $\sigma^2 S^2 dt$ in the previous Taylor expansion, and  
this simplifies into
\beq
&&\dt u + \demi\sigma^2F(\gamma)=0\label{main1},\\
&&F(\gamma) = \frac{\gamma}{1-\lambda \gamma},\label{main2}\\
&&\gamma=s^2\dss u.
\enq
{\it Remark.} Note that if we change assumption (\ref{fondambis2}) in 
\beq
\label{fondambis3} \text{Average execution price }& = & S(1+\lambda NS),
\enq 
then we get the equation
\ben
&&\partial_t u  + \demi \sigma^2\frac{\gamma}{(1-\lambda \gamma)^2} = 0,
\enn
as found in \cite{Liu}. This equation is not parabolic however.

When $\lambda \to 0$, we  we recover the standard Black-Scholes equation.
This equation clearly poses problems when $1-\lambda \gamma $ goes to 0, and even becomes negative.
This case arises when one has sold a convex payoff ($\gamma > 0$) and an initial move of $dS$ would be more than doubled during a naive re-hedging (see above) due to our market impact, $\lambda S^2 \Gamma dS \geq dS$.
Intuitively, the hedger runs after the spot, without being able to reach a point where he is hedged,and the spot runs away to infinity or to 0.
In that case, believing the model, when $1-\lambda \gamma <0$, , when the spot moves up, one should  {\bf sell} stocks instead of buying, because the market impact will make the spot go down.  One relies on the market impact to take back the spot at a level where we the portfolio is hedged. This situation is clearly nor realistic, nor acceptable from a trading perspective ...
On the other hand, if we consider a smooth function that satisfies for all time $t\in [0,T]$ the constraint
\ben \label{contrainteintro}
1-\lambda s^2 \partial_{ss} u(t,s) \leq 1  -\varepsilon \text{ for some } \varepsilon > 0,
\enn
solving (\ref{main1}, \ref{main2}) with terminal payoff $u(T,s)=\Phi(s)$ 
then our approach will be shown to be valid, and the  system (\ref{main1}, \ref{main2}) yields an exact replication strategy (we shall prove this in the verification theorem hereafter). Note that our approach has been presented in the case of a constant volatility, but would adapt with no modification to another volatility process, local or stochastic. The function $F$ is increasing, which guarantees that the time independent problem is elliptic, and thus the evolution problem is a priori well posed, although the question of the existence of solutions to the fully non-linear  pricing equation still remains.

Another (informal) way to see the constraint on $u$,  is to consider instead of $F$ 
\be\label{defF}
F(\gamma) &=& \frac{\gamma}{1-\lambda  \gamma} \hbox{ if } \lambda  \gamma < 1,\\
&=&+\infty \hbox{ otherwise}.
\en
The physical interpretation of the singular part is that areas with large positive $\gamma$ ( i.e such as $\lambda\gamma > 1$) will be quickly smoothed out and will instantaneously disappear, as if the final payoff was smoothed (again this argument will be made rigorous later on). This amounts to replace the solution $u$ by the smallest function greater than $u$ and satisfying the constraint ($\lambda s^2 \dss u \leq 1$) (a semi-concave envelope, the so-called "face-lifting" in \cite{SonTouz2}).
This is actually a common practice on derivatives trading desks: one replaces then a single call by a strip of calls, in order to cap the $\Gamma$, and this approach has been used by in \cite{BoLoZo2}.
It can be expressed by turning the system (\ref{main1}, \ref{main2}) into
\beq\label{main1bis}
&&\max\{\partial_t u  + \demi \sigma^2  F( \gamma ), \lambda\gamma - 1 + \varepsilon\} = 0,\\  
&&F(\gamma) = \frac{\gamma}{1-\lambda \gamma}\label{main2bis},
\enq
for some $\varepsilon>0$,
still with $\gamma = s^2\dss u$.
Under this formulation the problem enters  into the framework of viscosity solutions, see \cite{CrandallIshiiLions}.

Note that on the other hand,  areas with large negative $\gamma$ (the hedger is buying a convex payoff) would have very little diffusion, which also poses a problem as the equation is not uniformly parabolic any more.

\begin{figure}[ht]
\includegraphics[width=12cm]{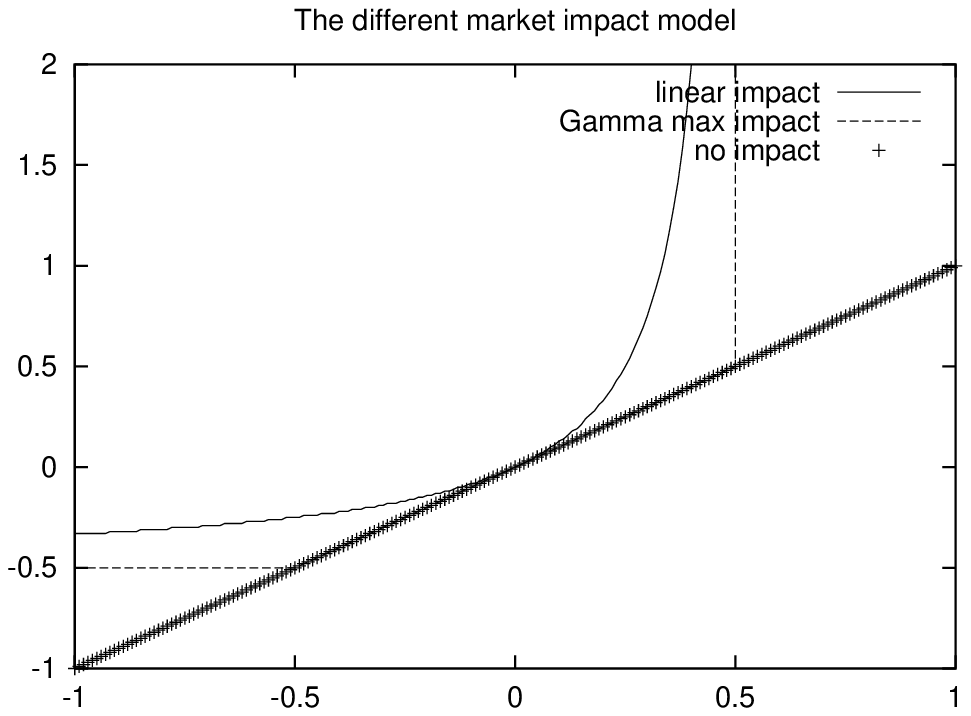}
\caption{The function $\Gamma_c \to F(\Gamma_c)$ used for the different models}
\label{t}
\end{figure}

 The functions $\Gamma \to F(\Gamma)$ are represented in Fig. \ref{t}. 

\section{Time Continuous Formulation of the Problem}

We now formulate our problem as a system of stochastic differential equations, as done in \cite{BoLoZo1}, \cite{BoLoZo2}. This formulation is similar to the one of Soner and Touzi \cite{SonTouz}, \cite{SonTouz2} of stochastic target problems, or to the formulation of backward stochastic  differential equations problems \cite{SonTouz4}. The crucial difference is that here the spot process $S_t$ itself has its dynamic modified by the controls.

We consider a probability space $(\Omega, \bF,\bP_0)$ supporting a standard Brownian motion $W_t^{\bP_0}$ and its associated filtration $({\cal F}_t)_{t\geq 0}$.  The {\it drift} $\nu_t$ will be a bounded adapted process, and we consider bounded  processes (the {\it controls}) $a_t, \Gamma_t$, adapted to the filtration ${\cal F}_t$. For two semi-martingales $u,v$, $\lb u,v\rb_t$ denotes their covariation (and $\lb u,u \rb_t$ denotes the quadratic variation of $u$).  Both $\sigma$ and $\lambda$ are given, they can be constant or have some dependency, which will be made explicit when needed. We will consider the following system of stochastic differential equations,
\beq
\frac{d S_t}{S_t} &=& \sigma dW_t^{\bP_0} + \lambda S_t d\delta_t + \nu' dt,\label{eds2}\\
d\delta_t&=&a_t dt + \Gamma_t dS_t\label{eds1},\\
dV_t&=& \delta d S_t + \demi \lambda S_t^2 d\lb \delta, \delta\rb_t,
\label{eds3}
\enq
where $S$ starts from $S_0$ at $t=0$.
It has been established rigorously in \cite{BoLoZo1}, Proposition 1, how to obtain this system as the limit of a discrete (in time) trading strategy. In particular, note the modified drift 
\beq\label{defnu'}
\nu'=\nu+ \Gamma s \sigma\ds(\lambda s^2).
\enq
At each time $t$ the hedger holds $\delta_t$ units of the risky asset, whose value is $S_t$.
From equations (\ref{eds2}, \ref{eds1}), $S_t$ and $\delta_t$ are two continuous semi-martingales.
Equation  (\ref{eds2}) states that the spot price is driven by an exogenous source of noise and by the market impact (note the non-trivial modification of the drift process $\nu$, however this will not affect the pricing equation).
Equation (\ref{eds3}) takes into account the order book's behaviour to give the value process of a trading strategy, it can be written as
\ben
dV_t = \delta_t dS_t + \demi\lambda \Gamma_t^2 S_t^2d\lb S,S\rb_t.
\enn
It is the continuous version of (\ref{dVheuristics}).

Combining (\ref{eds1}) and (\ref{eds2}) one sees right away that 
\ben
\frac{d S_t}{S_t}(1-\lambda S^2_t \Gamma_t) &=& \sigma_t dW_t^{\bP_0} + (\lambda_t S_t a_t + \nu_t') dt.
\enn
We will assume  for now the condition 
\beq\label{cond}
\text{ For some } \varepsilon > 0, 1-\lambda_t S^2_t \Gamma_t\geq \varepsilon.  
\enq
We have the first elementary result:
\begin{prop}
Let $a_t,\Gamma_t$ be bounded adapted processes. Assume that $\Gamma_t$ satisfies uniformly the condition (\ref{cond}).  Then there exists a unique strong solution to (\ref{eds1},\ref{eds2},\ref{eds3}).
\end{prop}
The {\bf replication problem}  is to find a self financed strategy, hence controls $a_t, \Gamma_t$, and an initial wealth $V_0$ such that
\ben
\Phi(S_T)= V_T, a.s.
\enn
On the other hand the {\bf super-replication} problem of \cite{SonTouz2} is to find the lowest initial wealth such that there exists an admissible control reaching the target, i.e.
\ben
\Phi(S_T) \leq V_T, a.s.
\enn
The two problems are not equivalent in all cases, in particular, if the associated HJB equation is not parabolic (see \cite{SonTouz3}). In that case the verification theorem does not apply, and the optimal strategy is not straightforward (and might not even be unique). As we will show, the pricing equation is in our case well posed, and by solving it we are able to exhibit directly an exact replication strategy.
We address the super-replication problem in two companion papers \cite{BoLoZo1}, \cite{BoLoZo2}.

For a smooth function $u(t,s)$ such that $u(T,\cdot) = \Phi$,  we consider the strategy given by $\delta(s,t)=\partial_s u(t,S_t)$ (and hence $\Gamma_t=\dss u(t,S_t)$). One obtains for the wealth
\ben
V_T&=&V_0+ \int_0^T \partial_su dS_t +\demi \int_0^T\lambda (\Gamma_t S_t)^2d\lb S,S\rb_t,
\enn
while It\^o's formula applied to  $u$ reads
\ben
u(T,S_T)&=&u(0,S_0) + \int_0^T \partial_su dS_t + \demi\partial_{ss}u d\lb S,S\rb_t + \dt u  dt.
\enn 
Obviously, a function $u(t,s)$ satisfying
\ben
\demi\dss u(t,S_t) \, d\lb S,S\rb_t + \dt u(t,S_t) \, dt=\demi \lambda (S_t\partial_{ss}u(t,S_t))^2\, d\lb S,S\rb_t
\enn
looks like the good candidate. 
We know that
\ben
\frac{dS_t}{S_t}=\frac{\sigma_t }{1-\lambda_tS_t^2\partial_{ss}u}dW_t^{\bP_0} + \tilde\nu_t dt 
\enn
for some  adapted process $\tilde\nu_t$ (that depends on $\nu$ and on $u$). Then the condition on $u$ turns into
\ben
\demi\partial_{ss}u\frac{\sigma^2s^2}{(1-\lambda s ^2\partial_{ss}u)^2} + \dt u =\demi \lambda (s \partial_{ss}u)^2\frac{\sigma^2s^2}{(1-\lambda s^2\partial_{ss}u)^2}.
\enn
\subsection{The pricing equation}
We rewrite the equation above as
\beq
&&\dt u + \demi \sigma^2F(\gamma)=0,\label{main-lambda-dep1}
\enq
with 
$F$ being given by
\beq
F(\gamma)=\frac{\gamma}{1-\lambda\gamma}, \gamma=s^2\dss u\label{main-lambda-dep2},
\enq
with  the terminal condition 
\beq
u(T,s)=\Phi(S).\label{main-lambda-dep3}
\enq

\subsection{The intensity dependent impact}

Let us now turn to a slightly more general heuristic, by allowing the parameter $\lambda$ to depend on the trading strategy. In the previous approach we have assumed that the market impact of buying $N$ stocks is, in terms of price, $\lambda NS^2$, regardless of the time on which the order is spread. We now assume that $\lambda = \lambda(s^2\Gamma)$. The quantity $\gamma=\Gamma s^2$ relates to the {\it trading intensity}, i.e. to the rate at which stocks are traded by the option's hedger, hence a gamma dependent market impact can make sense. The system of stochastic differential equations governing the evolution is now given by

\beq
\frac{d S_t}{S_t} &=& \sigma dW_t^{\bP_0} + \nu dt + \lambda(\gamma) S_t d\delta_t,\label{eds21}\\
d\delta_t &=& a_t+dt \Gamma dS_t \label{eds22},\\
dV_t &=& \delta_t dS_t + \demi \lambda(\gamma)S_t^2d \lb \delta,\delta\rb_t,\label{eds23}
\enq
still with $\gamma = \Gamma s^2$. 
Then the pricing equation remains (\ref{main-lambda-dep1}), but now $\lambda = \lambda(\gamma)$.
For the equation to be well posed, we need to have $$F(\gamma) = \frac{\gamma}{1-\lambda(\gamma)\gamma}$$
non-decreasing. 
We have $$F' = \frac{1+\lambda'\gamma^2}{(1-\lambda\gamma)^2},$$
hence the condition on $\lambda$ is that
\beq\label{ellip-lambda}
1+\lambda'\gamma^2>0.
\enq
This case is discussed in more details in the appendix, section \ref{genlambda}.

\subsection{The Verification Theorem}
We conclude this section by stating  a verification result and a representation formula for solutions to  the replication problem. We first introduce two versions of condition (\ref{contrainteintro}):
\beq\label{contrainte-large}
\lambda s^2\dss u &\leq & 1,
\enq
and the strict version of (\ref{contrainte-large})
\beq\label{contrainte-strict}
\lambda s^2\dss u &\leq & 1-\varepsilon \text{ for some } \varepsilon>0.
\enq

\subsubsection{The risk-neutral dynamic}
We consider on $(\Omega, \bF)$ a probability $\bP$ and $W^{\bP}$ a $\bP$-Brownian motion, and consider the solution of:
\beq
\frac{dS^{t,s}_r}{S^{t,s}_r}&=&\sigma^\gamma_t dW^{\bP}_t,\label{dSver1}\\
\sigma^\gamma_t&=&\frac{\sigma(t,S^{t,s}_r)}{1-\lambda\gamma(t,S^{t,s}_r)},\label{dSver2}\\
S^{t,s}_t&=&s,\label{dSver3}
\enq
When $t=0, s=S_0$ we might just write $S_r, r\geq 0$ instead of $S^{0,S_0}_r$.

\begin{theo}\label{theo_verif}
Let $u$ be a $C^3([0,T]\times \R^+)$ smooth solution of (\ref{main-lambda-dep1}, \ref{main-lambda-dep2}, \ref{main-lambda-dep3}) satisfying (\ref{contrainte-strict}). Assume also that 
\begin{itemize}
\item[i)] $\sigma, \sigma^{-1}, s\ds\sigma(t,s)$ are bounded,
\item[ii)] $\lambda, s\ds(\lambda(\gamma))$ are bounded.
\end{itemize}
Then there exists a  strong solution $(S_t,\delta_t, V_t)$ defined up time $T$
 to the system (\ref{eds21}, \ref{eds22}, \ref{eds23}) above,  such that $\delta_t = \ds u(t,S_t)$ and $V_t=u(t,S_t)$ and, ${\bP}_0-$almost surely,
\beq\label{verif}
V_T=u(0,S_0) + \int_0^T\partial_s u d S_t  
+ \demi \int_0^T \lambda\frac{\sigma^2\gamma^2}{(1-\lambda\gamma)^2} dt= \Phi(S_T).
\enq
The evolution of $S$ under $\bP_0$ is given by
\ben
\frac{dS_t}{S_t}&=&\frac{1}{1-\lambda\gamma} (\sigma dW_t^{\bP_0}+ \tilde \nu dt),\\
\tilde \nu &=& \nu'   -   \frac{\sigma^2\lambda^2\gamma}{(1-\lambda \gamma)^2}- \demi\ds(\sigma^2)\frac{\lambda s \gamma}{1-\lambda\gamma} - \demi\ds(\lambda(\gamma)) \frac{\lambda s \sigma^2\gamma^2}{(1-\lambda \gamma)^2},
\enn
with $\nu'$ defined in (\ref{defnu'}). There exists $\bP$ absolutely continuous with respect to $\bP_0$ such that $S$ satisfies (\ref{dSver1}, \ref{dSver2}, \ref{dSver3}) under $\bP$, 
and there holds
\beq\label{dualverif}
u(0,S_0) = \E^{\bP}\left(\Phi(S_T) - \frac{1}{2\lambda}\int_0^T(\sigma^\gamma(t,S_t)-\sigma(t,S_t))^2 dt\right).
\enq
In particular, the result of the Theorem hold if $\lambda$ is constant and $\Phi, \sigma$satisfy the assumptions i) to v) of Theorem \ref{ubvbm}.


\end{theo}

{\it Proof.} The proof of (\ref{verif}) is a simple application of It\^o's formula, since we restrict ourselves to the case where the solution of 
(\ref{main-lambda-dep1}) is smooth, the modified volatility $\sigma^\gamma$ remains bounded,  and thus the solution is a solution in the classical sense. For the expression of $\tilde\nu$, we just compute $d(\ds u(t,S_t))$. The conditions i) and ii) together with condition (\ref{cond}) imply that $\tilde \nu$ remains bounded. The condition ii) needs some a priori estimates to be enforced, but is trivially satisfied in the case where $\lambda$ is constant, which will be our main focus in this paper. Conversely the condition on $\sigma$ is straightforward to check, and  is also found to be necessary for the proof of uniform boundedness of the volatility (see Theorem \ref{ubvbm}).
Then under the assumptions on $u$, the existence of $\bP$ is a standard application of Girsanov's Theorem (see \cite{KaratzasShreve}). Under $\bP$, $S$ follows (\ref{dSver1}, \ref{dSver2}, \ref{dSver3}), and (\ref{dualverif}) follows  by observing that
\ben
\sigma^\gamma - \sigma = \sigma\frac{\lambda\gamma}{1-\lambda\gamma},
\enn
and then by taking the expectation of (\ref{verif}) under $\bP$.

$\hfill \Box$

This result shows that, under existence of a smooth solution $u$ to (\ref{main-lambda-dep1}), the claim $\Phi$ is replicable by the self financed strategy that consists in holding $\delta_t = \partial_su(t,S_t)$ stocks, and that the cost of this replication strategy is given for all time by $u(t,S_t)$.
Note again that the profit generated by the hedging strategy is no more the usual expression
$\int_0^T\partial_s ud S$, but includes an additional term due to the market impact (more exactly due to the difference of the price after market impact and the executed price, i.e. the liquidity costs). It is always positive (i.e. in favor of the option's seller). As observed above this seems surprising, but note that the change of volatility from $\sigma$ to $\sigma/(1-\lambda \gamma)$ acts always against the option's seller  , and the sum of the two impacts is always against the option's seller.

\section{Dual formulation of the problem}\label{dualsection}
We mention here the connection between our pricing equation and the dual formulation of second order target problems studied in \cite{SonTouzZhang}.
The equation we study here is still
\ben
&&\dt u  + \demi\sigma^2F(\gamma)=0,\\
&&F(\gamma)=\frac{\gamma}{1-\lambda \gamma},\\
&&u(T,s)=\Phi(s).
\enn
$\gamma=s^2\dss u$, and we assume that $\Phi$, $\sigma$ satisfy the conditions of Theorem \ref{theo-main}, so that $u$ is smooth on $[0,T]\times \R_+^*$. $F$ is convex, and we compute the $\sigma-$Legendre transform (a slightly modified Legendre transform) of $F$:
\beq
\label{defFstar}F_\sigma^*(a)&=&\demi\sup_\gamma\{a\gamma - \sigma^2F(\gamma)\}\\
&=&\frac{1}{2\lambda}(a^{1/2}-\sigma)^2.\nonumber
\enq
Note that from the strict convexity of $F$, there will hold
\beq\label{defFa}
\demi\sigma^2F(\gamma)=\sup_{a}\left\{\demi a\gamma - F^*_\sigma(a)\right\},
\enq
and that the supremum in (\ref{defFa}) is reached for $a=\hat{a}=(\sigma^\gamma)^2=\frac{\sigma^2}{(1-\lambda\gamma)^2}$, which we recognize as the modified variance in our model.
For $W^\bP$ a $\bP-$ Brownian motion, with filtration $({\cal F}_t)_{t\geq 0}$, 
we define
\beq{\cal A}_T=&& \Big\{a_{t, t\in [0,T]}, a \text{ is } {\cal F_t}-\text{predictable},\label{defA}\\ 
&&\exists \bar{a}\in R_+^* \text{ such that }0 \leq a_t \leq \bar{a}\;\bP-a.s. \Big\}\nonumber
\enq
 
For $a\in {\cal A}_T$ we can define $S^{a,t,s}$ such that
\beq\label{defsa1}
S^{a,t,s}_{t}&=&s,\\
dS^{a,t,s}_{r}&=&S^{a,t,s}_{r} (a_r)^{1/2} dW^{\bP}_r, t\leq r \leq T,\label{defsa2}
\enq
and define, for $\Phi$ a terminal condition,
\ben
u^a(t,s) = \E^{\bP_0}\left\{\Phi(S^{a,t,s}_T) - \int_0^T F_\sigma^*(a_u)du\right\}.
\enn
Then applying It\^o's formula to $u(r,S^{a,t,s}_r)$ (which is allowed from the regularity of $u$) we have that
\ben
\Phi(S^{a,t,s}_T) &=& u(t,s) + \int_t^T\left(\dt u + \demi a_r (S^{a,t,s}_r) \dss u\right)dr + \int_t^T\ds u \, a_r^{1/2} S^{a,t,s}_rdW^{\bP}_r \\
&= & u(t,s) + \int_t^T\left(\demi a_r (S^{a,t,s}_r) \dss u - \frac{\sigma^2}{2}
F((S^{a,t,s}_r)^2 \dss u)\right)dr\\
&& + \int_t^T\ds u\,  a_r^{1/2} S^{a,t,s}_rdW^{\bP}_r,
\enn
hence from (\ref{defFstar}),
\ben
\E^{\bP}\left(\Phi(S^{a,t,s}_T) \right) \leq u(t,s) +  \E^{\bP}\left( \int_0^T F_\sigma^*(a_r)dr\right),
\enn
which shows that $u^a(t,s) \leq u(t,s)$, with equality if and only if $a=\hat{a}=\sigma^\gamma$, hence $S^{\hat{a}}$ is the risk-neutral diffusion (\ref{dSver1}, \ref{dSver2}, \ref{dSver3}).
%
%
%
Letting, for $a\in {\cal A}_T$ 
\beq\label{defC}
{\cal C}_\sigma(a) = \int_0^T (a_t^{1/2}-\sigma(t',s))^2dt',
\enq
we have thus obtained the following:
\begin{theo}
let  $u$ be a  $C^{1,3}([0,T]\times (0,+\infty))$ smooth solution to (\ref{main-lambda-dep1}, \ref{main-lambda-dep2}, \ref{main-lambda-dep3}) satisfying (\ref{contrainte-strict}), then
\beq\label{suprep}
u(t,s)=\sup_{a \in {\cal A_T}}\Big\{ \E^{\bP}\{\Phi(S^{a,t,s}_T) - \frac{1}{2\lambda}{\cal C}_\sigma(a)\}\Big\},
\enq
and the supremum is attained for $\hat{a}=\sigma^\gamma$ as in (\ref{dSver2}).
In particular this holds true if $\Phi, \sigma$ satisfy the assumptions to Theorem \ref{theo-main} for global regularity on $[0,T]$. 
\end{theo}

{\it Remark.} In Theorem \ref{local-rep-2}, our result is stronger, as formula (\ref{suprep}) holds for any $\Psi \leq \Phi$ such that $(1-\lambda s^2\dss \Phi)(\Phi-\Psi)=0$, and hold up to cases where $E^{\bP}(\Phi(S_T))$ might no be finite.
We will also show that the optimal process $S_t$ remains a true martingale up to time $T$ under mild assumptions on $\Phi$, which allow for the optimal $\hat{a}$ to be unbounded, hence $\hat{a}\notin {\cal A}_T$.

\subsection{Relation to a problem of optimal transport} As mentioned in the introduction, having found $u$ by solving (\ref{main-lambda-dep1}), and specifying an initial distribution ${\cal L}_0$ for $S_0$, 
we let ${\cal L}_T^\Phi$ be the law of $S_T^{\hat{a},0,S_0}$. Then $\hat{a}=\frac{\sigma^2}{(1-\lambda\gamma)^2}$ realizes also
\ben
\hat{a}=\text{argmin}\Big\{\E^{\bP}{\cal C}_\sigma(a)\Big\}
\enn
where the infimum is taken over all the process $S^{a,0,{\cal L}_0}, a\in {\cal A}_T$, that have laws ${\cal L}_0$ at time 0, ${\cal L}^\Phi_T$ at time $T$, and follow (\ref{defsa2}). 
This problem of optimal transport by martingales has been studied by Touzi and Tan in \cite{TanTouz}. 
Therefore, our regularity results can be seen as a step towards the regularity of optimal transport by diffusion, for this particular cost. Moreover, as we will see in Theorem \ref{maintheo}, when the volatility is constant, we obtain a closed formula to express the final density in terms of the terminal condition $\Phi$.

\subsection{Interpretation in terms of robust hedging} Assuming one is trying to find a robust hedging strategy for the claim $\Phi$ when the volatility of the underlying is unknown. The formula (\ref{suprep}) gives the optimal upper bound on the price of the claim $\Phi$ over all possible diffusions that satisfy 
\ben
\E^{\bP}{\cal C}_\sigma(a)\leq \E^{\bP}{\cal C}_\sigma(\hat{a}),
\enn
as it gives the cheapest hedging strategy that will super-replicate the claim $\Phi$ for $a\in {\cal A}_T$ satisfying the above constraint.

Note also that the formula (\ref{defC}) can be interpreted as the payoff of a volatility derivative, and that (\ref{suprep}) gives the cheapest way to super-replicate the claim $\Phi$ by buying $\frac{1}{2\lambda}$ units of the ${\cal C}_\sigma$ and delta-hedging, or alternatively, a lower bound on the price of the derivative ${\cal C}_\sigma$ given the price of the claim $\Phi$.

\section{Smooth solutions via Legendre-Fenchel Transform}

In this section we prove existence and regularity of the solution to (\ref{main-lambda-dep1}, \ref{main-lambda-dep2}, \ref{main-lambda-dep3}), in the cases of constant or local volatility $\sigma(t,s)$, and with constant market impact parameter $\lambda$. In particular, our result will give the conditions on $\sigma$ and $\Phi$ for which solution $u$ has enough regularity to satisfy  the assumptions of the Verification Theorem (Theorem \ref{theo_verif}), and to define the dynamics of $S_t$.
The case of non-constant $\lambda$ is treated in the appendix.
\subsection{Notations}

\begin{itemize} 
\item[-] As we will work with H\"older spaces, we note for $A\subset [0,T], B\subset \R)$, $k,m\in \N$, $0<\alpha, \beta<1$, $\|u\|_{C^{k+\alpha, m+\beta}_{s,t}(A\times B)}$ (resp. $\|u\|_{C^{k+\alpha}_{s}(A\times B)}$) (resp. $\|u\|_{C^{m+\beta}_{t}(A\times B)}$) the usual H\"older norm of $u$ of order $k+\alpha$ with respect to $s$ and order $m+\beta$ with respect to $t$. When no subscript $t$ or $s$ is specified, the  continuity will be with respect to $s$.

\item[-] We shall denote $C^{k+\alpha, m+\beta}_{s,t,loc}(A\times B)$ the space of functions with bounded $C^{k+\alpha, m+\beta}_{s,t}(K)$ norm for all compact sets $K\subset A\times B$.

\item [-] Classically we will denote $\R_+^*=(0,+\infty)$.

\item[-] Whenever needed, we will consider a Brownian motion $W^\bP_t$ supported on $(\Omega,\bF, \bP, ({\cal F}_t)_{t\geq 0})$ a filtered probability space, and denote $\E^{\bP}$ the expectation under the probability measure $\bP$.

\end{itemize}

\begin{defi}\label{def-classical}
We shall say that $u$ is a classical solution to (\ref{main-lambda-dep1}, \ref{main-lambda-dep2}, \ref{main-lambda-dep3}) if $u\in C^{2,1}_{s,t,loc}([0,T)\times \R_+^*)\cap C^0_{t,s}([0,T]\times \R_+^*)$, satisfies (\ref{main-lambda-dep1}) on $[0,T)\times \R_+^*$, (\ref{main-lambda-dep3}), and if 
\ben
1-\lambda s^2\dss u < 1
\enn
 on  $[0,T)\times \R_+^*$.
\end{defi}

\subsection{Some facts about the Legendre-Fenchel transform}
A reference on this topic is \cite{R}.
\begin{defi}\label{defilegendre} Let $u:\R\to \R\cup\{+\infty\}$, its Legendre transform is defined by 
\beq\label{deflegendre}
u^*(y)=\sup\{xy - u(x), x\in \R\}.
\enq
If $u$ is convex and l.s.c. then $(u^*)^*=u$. 
Moreover 
\begin{itemize}
\item[-] If $u$ is continuously  differentiable at $x$ then, 
\ben
u^*(\dx u (x)) + u(x) &=& \dx u(x)\cdot x,\\
\dy u^*(\dx u(x))&=&x,
\enn
and the reverse equality hold since $(u^*)^*=u$.
\item[-]
If $y \notin \overline{\dx u(\R)}$ then $u^*(y)=+\infty$.
\item[-] 
At a point where $\dxx u(x)$ is defined and positive,  $\dyy u^*(\dx u)$ is defined and satisfies
\ben
\dxx u(x) \dyy u^*(\dx u) =1.
\enn
\item[-] If $u$ depends smoothly on a parameter $t$, for all $x\in\text{Dom}(u)$,
\ben
\dt u^*(t,\dx u) + \dt u(t,x) = 0.
\enn
\end{itemize}
\end{defi}

\subsection{Transformation of the  pricing equation via Legendre transforms}
Starting from a classical (see Definition \ref{def-classical}) solution of (\ref{main-lambda-dep1}, \ref{main-lambda-dep2}, \ref{main-lambda-dep3}), we consider 
\beq\label{defv}
v = -\lambda u(t,s) - \ln(s)-1 \text{ if }s>0, +\infty \text{ otherwise}.
\enq
Then $v$ is convex under (\ref{contrainte-large}), and satisfies
\beq\label{eqonv}
\dt v -\frac{\sigma^2(t,s)}{2} \frac{1}{s^2\dss v }=-\frac{\sigma^2}{2}.
\enq
Consider $v^*$  the Legendre transform of $v$ defined by
\beq\label{defv*}
v^*(y) = \sup \{ sy - v(s), s>0 \}.
\enq
From Definition \ref{defilegendre}, for the pair ($y(t,s),s(t,y)$) where the maximum is attained, there will hold
\ben
v(t,s)+v^*(t,y)=sy,\\
y(t,s)=\ds v(t,s),\\
s(t,y)= \dy v^*(t,y),\\
\dss v(t,s) \dyy v^*(t,y)=1.
\enn
Moreover, one will have
\ben
v^*(t,y(t,s))&=&\ln(s) + \lambda(u -s\ds u),\\
\dt v(t,s) &+& \dt v^*(t,y(t,s))=0.
\enn
Then it follows that
\beq
\dt v^* + \frac{\sigma^2(t,s)}{2}\frac{\partial_{yy}v^*}{(\partial_y v^*)^2}&=&\frac{\sigma^2(t,s)}{2},\label{eqonv*0}\\
s&=& \dy v^*(t,y).\nonumber
\enq
By straightforward computations, $w$, the inverse function of $v^*$, satisfies
\beq\label{linw}
\dt w + \frac{\sigma^2(t,s)}{2}(\dxx w + \dx w)&=&0,\\
s&=&\frac{1}{\dx w(t,x)}.\nonumber
\enq
To construct properly $w$, we use also the Legendre transform. Since $v^*$ is increasing, a primitive of $v^*$ is convex, hence one will have for $W=\int v^*$, $\dx W^*(v^*(y))=y$ at any point where $v^*$ is continuous, i.e. everywhere in the domain of $v^*$.

\subsubsection*{Construction of the terminal value for the transformed equations}

\begin{prop}\label{definverse}
Let $\Phi:\R_+^* \to \R$ satisfy $\lambda s^2\dss \Phi \leq 1$.
Define $v_T, v^*_T, w_T$ such that
\ben
v_T &=& - \ln(s) - \lambda \Phi -1 \text{ if }s>0, +\infty \text{ otherwise}, \\
v^*_T &=& (v_T)^*,\\
w_T&=&[v^*_T]^{-1}\text{ defined as } w_T=\dx\left(\int v^*_T\right)^*,\\
V_T(s) &=& v^*_T(\ds v_T),\\
\cS_T &=& \dy v^*_T(w_T))=\frac{1}{\dx w_T},
\enn
Then
\begin{enumerate}
\item $v^*_T$ is non decreasing ,
\item $v^*_T(\ds v_T)= \ln(s)+\lambda(\Phi - s\ds\Phi)$ wherever $\ds v_T$ exists,
\item $\lim_{y\to -\infty} \dy v^* = 0$, 
\item either $\exists L \in \R$ such that  $v^* \equiv +\infty$ above $L$
or $\lim_{y\to +\infty}\dy v^* = +\infty$.
\item If $\lim_{s\to 0} v_T$ is finite then $w_T$ is identically  $-\infty$ below this limit, otherwise $w_T$ is finite everywhere and $\lim_{x\to -\infty} \dx w_T = +\infty$.
\item If $\ds v_T$ is constant on some interval $(C,+\infty)$ then $\dx w_T$ is identically  $0$ above $C'$ for some $C'$, otherwise $\dx w_T$ is positive everywhere and $\lim_{x\to +\infty} \dx w_T = 0$.
\item For all $s$ where $\lambda s^2 \dss \Phi < 1$ and $\ds\Phi$ is continuous there holds
\ben\label{defS}
\cS_T(v^*_T(\ds v_T(s)))=\cS_T( \ln(s)+\lambda(\Phi - s\ds\Phi))= s.
\enn
\item Under assumption (\ref{kernel}), one has necessarily that
\ben
\lim_{s\to 0} v_T &=& +\infty,\\
\lim_{y\to -\infty} v_T^* &=& -\infty.
\enn
\end{enumerate}
From (5) and (8) we thus have Lebesgue a.e.
\beq
\label{defVphi}
V_T(s)&=&\ln(s)+\lambda(\Phi-s\ds\Phi),\\
\cS_T(V_T(s))&=&s.\label{defcS}
\enq

\end{prop}

{\it Proof.} 
\begin{enumerate}
\item The first point comes from the fact that $v_T=+\infty$ for $s<0$. 

\item This is just the definition of the Legendre transform, and the fact that $y=\ds v_T$ for the optimal  $s$ (see Definition \ref{defilegendre}).

\item The third point comes the fact that $v$ is defined and finite on $(0, +\infty)$. If one had $\lim_{y\to -\infty}\dy v^* = l >0$ then this would imply that $v\equiv +\infty$ below $l$.

\item The two cases correspond to $\lim_{s\to +\infty}\ds v$ being either finite (equal to $L$) of $+\infty$.

\item $v_T=+\infty$ below 0 implies that $\lim_{y\to -\infty} v_T^*= - \lim_{s\to 0} v_T$. If $\lim_{s\to 0} v_T(s)$ is finite, then $\lim_{y\to -\infty} v^*= -\lim_{s\to 0} v_T(s)$. This in turn implies that $w_T=-\infty$ below this limit. The other assertion follows from point 3 since $\dx w_T=\frac{1}{\dy v^*}$.

\item If $\ds v=l$ above some value $s_0$, then  $\lim_{y\to l}\dy v^*_T=s_0$ and $v^*_T=+\infty$ above $l$. Then $w_T$ is constant above $v^*_T(l)$.

\item Point 7 is just the fact that $[w_T]^{-1} = v^*$, hence 
\ben
\dx w_T(v^*_T(y)) = \frac{1}{\dy v^*_T}(y),
\enn
and the assertion follows.

\item For point 8, the first part follows from point 5, and implies the second part. Note that from (\ref{kernel}), $\dx w_T$ and $w_T$ are finite everywhere. 
\end{enumerate}

$\hfill \Box$

\begin{figure}[ht]
\includegraphics[width=12cm]{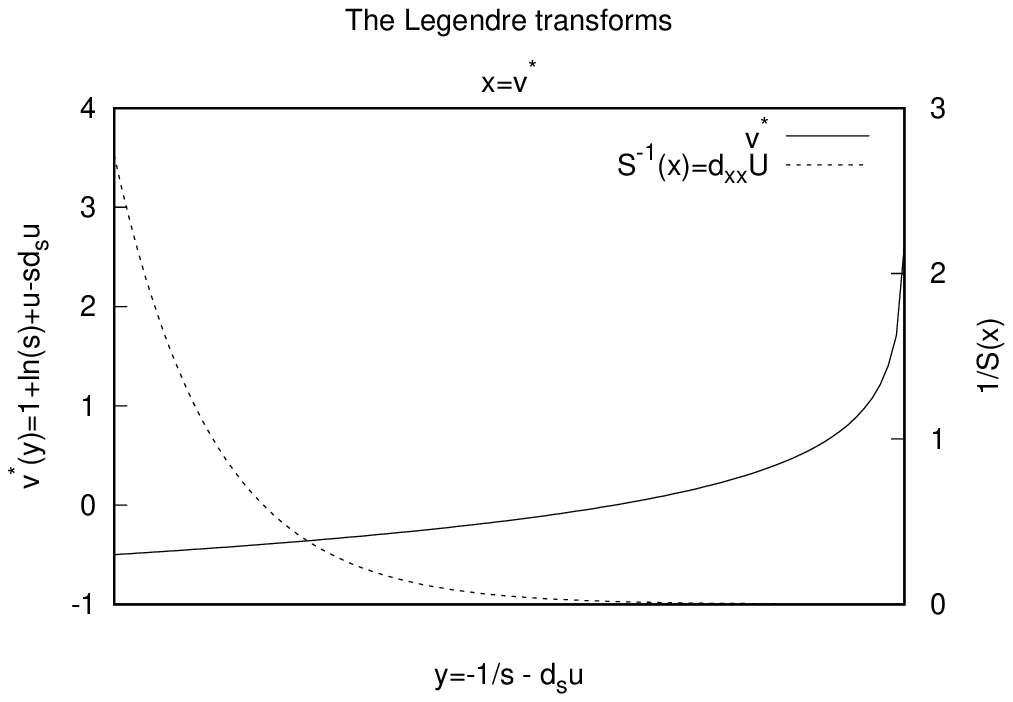}
\caption{$v^*$ and $\cS_T$ for $\Phi(s)=s-\demi\ln(s), \lambda=1$}
\label{legendre2}
\end{figure}

\newcommand{\tu}{\tilde{u}}
\newcommand{\tv}{\tilde{v}}

\subsection{Main assumptions}
\begin{defi}
$\Phi:\R_+^* \to  \R$ satisfies (\ref{kernel}) with parameter $\sigma$ if
\beq\label{kernel}
\int_{\R} \exp\Big(-\frac{x^2}{2\sigma^2(T+\varepsilon)}\Big)|\Phi(x)|dx \text{ is finite for some } \varepsilon >0.
\enq
When $\sigma$ depends on $(t,s)$ we introduce the following condition
\beq
\label{condsig}
\forall t\in [0,T], s>0, \usig \leq \sigma(t,s) \leq \bsig,
\enq
for positive constants $\bsig, \usig$,
and $\sigma$ will then be said to satisfy (\ref{condsig}) with parameters $\usig, \bsig$

\end{defi}

\subsection{The case of constant volatility}

We now show that (\ref{main-lambda-dep1}, \ref{main-lambda-dep2}, \ref{main-lambda-dep3}) admits a unique smooth solution when $\sigma, \lambda$ are constant. In this case $\dx w$ solves also (\ref{linw}), hence one can recover the function $v^*=\ln(s)+\lambda(u-s\ds u) \to \frac{1}{s}$ by solving a simple heat equation, which then leads to the solution $u$. Therefore, the condition for existence of a smooth solution can be stated as a condition on the function $\cS_T: v^* \to s$ at time $T$. 
We will have the following result:
\begin{theo}\label{theocstvol}
Let $\Phi$ satisfy  $\lambda s^2\dss \Phi \leq 1$.
Let $V_T=\ln(s)+ \lambda(\Phi- s\ds \Phi)$, and let $\cS_T:\R\to [0,+\infty]$ be the inverse of $V_T$ as constructed in  Proposition \ref{definverse}, (\ref{defVphi}, \ref{defcS}).
Assume that $\frac{1}{\cS_T}$ satisfies (\ref{kernel}) with parameter $\sigma$
Then 
\begin{itemize}
\item[i)]
There exists a unique $u$ classical solution to (\ref{main-lambda-dep1}, \ref{main-lambda-dep2}, \ref{main-lambda-dep3}). It belongs to $C^\infty([0,T)\times \R^*_+)$. 

\item[ii)]For all $(t,s)\in [0,T]\times\R_+^*$,
\ben
\inf_s \{s^2\dss\Phi\} \leq s^2\dss u(t,s) \leq \sup_s \{s^2\dss \Phi\}.
\enn

\item[iii)]
If for $k\geq 2, \nu>0$, $\Phi\in C^k(\R^*_+)$ and $\lambda s^2\dss \Phi \leq 1-\nu$, then 
\ben
|s^k\partial^k_su|_{\Linf([0,T]\times\R_+^*)} &\leq& C_k\left(|s^{k'}\partial^{k'}_s\Phi|_{\Linf(\R_+^*)}, k'\leq k\right).
\enn
\end{itemize}

\end{theo}

{\it Remark.} As we will see, $\cS_T$ is positive and increasing, thus  the condition (\ref{kernel}) on $\Phi$ is only about the behaviour of $\Phi$ near 0. It allows  the set $\{1-\lambda s^2\dss \Phi =0\}$ to be non-empty, in particular any globally Lipschitz function $\Phi$ such that 
\ben
\lambda s^2\dss\Phi \leq 1
\enn
satisfies (\ref{kernel}).
Surprisingly, even if $\lambda s^2\dss \Phi \equiv 1$ for $s$ above a certain threshold, condition (\ref{kernel}) might be satisfied, as it just implies that $\frac{1}{\cS_T}$ is identically 0 above a certain threshold. Conversely if $\lambda s^2\dss \Phi \equiv 1$ for $s$ close to $0$ then (\ref{kernel}) can not be satisfied. This is a feature of the log-normal dynamics that "send the mass to 0".

{\it Proof.} Starting from the terminal payoff  $\Phi$, as explained above, one constructs $v_T$, and then $v^*_T$, $w_T$ and then $\frac{1}{\cS_T}=\dx w_T$ which is defined almost everywhere on the set $\dy v^*_T >0$. For $t<T$, let $w$ be given by
\beq\label{defH}
w(t,x)=\frac{1}{\sigma(2\pi(T-t))^{1/2}}  \int_{\R} \exp\Big(-\frac{(x+\frac{\sigma^2}{2}(T-t)-z)^2}{2\sigma^2(T-t)}\Big)w_T(z)dz.
\enq
Condition (\ref{kernel}) implies that $w$ is well defined, $w$ solves \ref{linw} (which is a heat equation with constant coefficients), and so does $\dx w$,  and the properties listed in Proposition \ref{definverse} ensure that $w$ belongs to $C^\infty_{loc}([0,T)\times \R)$ and for $t<T$, $w$ is strictly increasing and strictly concave, $\dx w$ being given by
\beq\label{defh}
\dx w(t,x)&&=\\
&&\frac{1}{\sigma(2\pi(T-t))^{1/2}}  \int_{\R} \exp\Big(-\frac{(x+\frac{\sigma^2}{2}(T-t)-z)^2}{2\sigma^2(T-t)}\Big)\frac{1}{\cS_T(z)}dz,\nonumber
\enq
and having limits $+\infty$ at $-\infty$ and $0$ at $+\infty$.

Then one follows backward our previous transformations of $u$:
\begin{itemize}
\item[-] The inverse of $w^{-1}$ (in the sense of inverse functions) will satisfy (\ref{eqonv*0}), \item[-] its Legendre transform $(w^{-1})^*$ will then satisfy (\ref{eqonv}), 
\item[-] finally $u(t,s)= -\lambda^{-1}((w^{-1})^*(t,s)+\ln(s))$ will satisfy (\ref{main-lambda-dep1}, \ref{main-lambda-dep2}, \ref{main-lambda-dep3}).
\end{itemize}

The bounds on $s^2\dss u$ are a direct consequence of the following lemma:

\begin{lemme}\label{max-princ-dssu}
Let $u$ be the classical solution to (\ref{main-lambda-dep1}) on $[0,T)$ with constant $\sigma$, constructed as above from $w$. Then $\sup_{s>0} \{\gamma(t,s)\}$ is non-increasing and $\inf_{s>0}\{\gamma(t,s)\}$ is non decreasing.
\end{lemme}
{\it Proof.} Observe that for $u$ a classical solution of (\ref{main-lambda-dep1}), and $w$ defined from $u$, one has $\dx w= \frac{1}{\cS}$ and $\dxx w = -\frac{1}{\cS(1-\lambda \gamma)}$. We will have proved the lemma if we show that 
for any $C>0$, if $C\dx w_T + \dxx w_T \geq 0$ (resp. $\leq 0$) then for time $t<T$ the same inequality holds. This will hold if  $C\dx w_T + \dxx w_T$ which solves the heat equation, satisfies the maximum principle. This type of result is in general is not true on the whole line without any growth assumptions, see \cite{Tycho}, but in our case this is granted as we have constructed $w$ through the representation formula (\ref{defh}).

$\hfill \Box$

Bounds on $s^k\partial^k_s u$ can be obtained obtained by looking at $v(t,y)=u(t,e^y)$, that solves
\ben
\dt v + \frac{\sigma^2}{2}\frac{\dyy v -\dy v}{1-\lambda (\dyy v - \dy v)}=0.
\enn

The bounds on $\partial^k_y v$ are classically obtained by differentiating the equation, and imply the bounds on $s^k\partial^k_s u$.

Finally the uniqueness is a consequence of Widder's Theorem (see \cite{KaratzasShreve}, Chapter 4, Theorem 3.6) since starting from $u$ a classical solution to (\ref{main-lambda-dep1}) one can build $w$ from $u$ as explained above, and then $\dx w$ will be a positive solution to the heat equation, for which uniqueness holds.

$\hfill \Box$

\subsection{The case of non-constant volatility}
Here we show the following a priori estimate
\begin{theo}\label{ubvbm}
Assume that there exist constants $M,A,B,\varepsilon_1, \varepsilon_2,C_\sigma$ with $M, \varepsilon_1, \varepsilon_2$ positive, such that, for $\Lambda=\lambda^{-1}$:
\begin{itemize}
\item[i)] For $s\leq 1/M$, 
\ben (\varepsilon_1 - \Lambda)\ln(s) -A + Bs \leq \Phi(s) \leq (\varepsilon_1 - \Lambda)\ln(s) + A+  Bs.
\enn
\item[ii)] For $s\geq M$,
\ben (\varepsilon_2 - \Lambda)\ln(s) -A +Bs  \leq \Phi(s) \leq (\varepsilon_2 - \Lambda)\ln(s) + A + Bs.
\enn
\item[iii)] $\underline{\sigma}  \leq \sigma(t,s)\leq \bar\sigma$.
\item[iv)] $|s\ds \sigma|+ |s^2\dss \sigma|+|\dt\sigma|+|s\partial_{st}\sigma| \leq C_\sigma$ .
\end{itemize}
Let $u$ be a classical solution to (\ref{main-lambda-dep1}, \ref{main-lambda-dep2}, \ref{main-lambda-dep3}), then for all $\tau >0$ small there exists $C_\tau(M,A,B,\varepsilon_1, \varepsilon_2,C_\sigma)$ such that for $0\leq t\leq T-\tau$, $\forall s>0$,

\beq\label{s2u}
-C_\tau \leq s^2\dss u(t,s) &\leq & \Lambda -1/C_\tau,\\
\label{s3u}
|s^3 \partial^3_s u| &\leq & C_{\tau}.
\enq
If moreover
\begin{itemize}
\item[v)] $\Phi\in C^{3+\alpha}$, $s^3\partial^3_s\Phi$ is bounded and $\Phi$  satisfies (\ref{s2u}),  
\end{itemize}
then $u$ satisfies (\ref{s2u}, \ref{s3u}) on $[0,T]\times \R_+^*$.

\end{theo}

{\it Remarks.} 
The interest of this result is to be able to define the dynamics of $S_t$ up to time $T-\tau$ for all $\tau>0$ without relying on stopping times. For this we need a uniform bound in space like (\ref{s2u}).

The conditions on $\Phi$ might not be minimal, but they allow for a large class of payoffs: all payoffs with linear or logarithmic growth at 0 and infinity.  In particular any combination of vanilla options.

If $\varepsilon_1=\varepsilon_2$ the proof becomes much simpler. The interest of the result lies in the fact that one can prescribe independently the behaviour at $0$ and $+\infty$.

{\it Proof of Theorem \ref{ubvbm}.} 

We recall that
\ben
\dt v^* + \frac{\sigma^2(t,s)}{2}\frac{\partial_{yy}v^*}{(\partial_y v^*)^2}=\frac{\sigma^2(t,s)}{2},
\enn
and using $s=\dy v^*$ this yields
\beq\label{eqonv*}
\dt v^* + \frac{\sigma^2(t,\dy v^*)}{2}\frac{\partial_{yy}v^*}{(\partial_y v^*)^2}=\frac{\sigma^2(t,\dy v^*)}{2},
\enq
while, as noted in (\ref{linw}) the inverse of $v^*$, $w$, follows
\beq\label{eqonw}
\dt w +  \frac{\sigma^2(t,1/\dx w)}{2}(\dxx w +\dx w)=0.
\enq
We will treat (\ref{eqonv*}) as an equation of the general form
\beq\label{parabogene}
\dt v^* + A(t,\dy v^*) \dyy v^* + B(t,\dy v^*) =0.
\enq
The strategy will be the following:
\begin{itemize}
\item[-] For equation (\ref{parabogene}), there exists (see \cite{Lieberman})  local regularity results that yield $C^{1,\alpha}$ regularity in space conditional to uniform ellipticity of $A$ and Lipschitz a priori estimates on $v^*$.
\item[-] In order to have those a priori estimates, we will use a barrier argument: we will show by the comparison principle that the solution is pinched between an upper and a lower bound, and this control used with the convexity of the solution will in turn lead to a control of the  gradient.
\item[-] The H\"older regularity of $\dy v^*$ will then imply that (\ref{eqonv*}) can be looked at as a linear uniformly parabolic equation, with H\"older continuous coefficients, and this in turn through Schauder estimates (see \cite{Lieberman} again) leads to $C^{1+\alpha/2, 2+\alpha}_{t,y}$ regularity.
\item[-] The growth conditions i) and ii) will then allow by a scaling argument to show the uniform bound (\ref{s2u}).
\end{itemize}  
 Note that from our assumption $\lim_{s\to +\infty}\ds\Phi(s)=B$, hence $v^*(T,y)\equiv +\infty$ for $y > -\lambda B$, and the equation (\ref{eqonv*}) has a singular boundary condition which will need careful treatment. On the other hand, from points 3 and 8 of Proposition \ref{definverse}, $\lim_{y\to -\infty}\dy v^*(t,y)=0$  and $\lim_{y\to -\infty} v^*(t,y)=-\infty$. One should  think of $v^*$ as a perturbation of $y\to -\ln(-\lambda B -y)$, see Fig. \ref{legendre2}. 

We will need for our estimates the following comparison result:
\begin{lemme}\label{lemmedur}
Let $w_T$ be concave and non-decreasing.
Assume that $\usig \leq \sigma(t,s)\leq \bsig$ and 
let $\bar{w}, \underline{w}$ be defined by 
\beq\label{defbarw}
\bar{w}(t,x) &=& \frac{1}{\bsig\sqrt{2\pi(T-t)}}\int_\R \exp\left(-\frac{(x+\demi \usig^2 (T-t)-z)^2}{2\bar{\sigma}^2(T-t)}\right) w_T(z)dz,\\
\uw(t,x) &=& \frac{1}{\usig\sqrt{2\pi(T-t)}}\int_\R \exp\left(-\frac{(x+\demi \bsig^2 (T-t)-z)^2}{2\usig^2(T-t)}\right) w_T(z)dz.\label{defuw}
\enq
Let $w$ be a $C^{2,1}_{x,t,loc}$ classical solution to 
\ben
\dt w + \demi\sigma^2(t,1/\dx w)(\dxx w + \dx w)=0
\enn
with $\dxx w \leq 0$,$\dx w>0$ on $[0,T)$. Then
\beq
\bw \leq w \leq \uw.
\enq
In particular if $u$ is a classical solution to (\ref{main-lambda-dep1}, \ref{main-lambda-dep2}, \ref{main-lambda-dep3}) on $[0,T)$, and $w$ is obtained from $u$ by the above procedure, then the conclusion holds true.
\end{lemme}

The proof is deferred to the appendix, section \ref{prooflemmedur}. It is a simple comparison principle, but is not trivial, since following some famous counterexamples by Tychonoff, (see \cite{Tycho})  uniqueness and comparison for solutions of the heat equation on the whole line does not hold unless some growth conditions are imposed. Here we do not need any growth but we use the concavity of the solution, hence this Theorem can be seen as a Widder's type Theorem which states uniqueness of positive solutions to the heat equation.

We next need the following lemma:
\begin{lemme} \label{finebouds}
Under assumptions of Theorem \ref{ubvbm}, one can find another constant $A'$ instead of $A$ such that the properties \textrm{i)} and \textrm{ii)} are satisfied by the solution $u$ on $[0,T]$.
\end{lemme}

{\it Proof of Lemma \ref{finebouds}.} By direct computations, the assumptions on $\Phi$ imply for $v^*(T)$ that, for some constants $C,D$:
\begin{itemize}
\item[i)] As $y$ goes to $-\infty$,
\ben 
-\lambda\varepsilon_1 \ln(-(y+\lambda B))  -\lambda C  \leq v^*(T) \leq-\lambda \varepsilon_1 \ln(-(y+\lambda B))  +\lambda C.
\enn
\item[ii)] As $y$ goes to $-\lambda B^-$,
\ben 
-\lambda\varepsilon_2 \ln(-(y+\lambda B))  -\lambda D  \leq v^*(T) \leq-\lambda \varepsilon_2 \ln(-(y+\lambda B))  +\lambda D.
\enn
\end{itemize}
We start  by constructing the barriers $\bv, \uv$ that are the inverse functions of $\bw, \uw$ in Lemma \ref{lemmedur}. Lemma \ref{lemmedur} implies then that 
\ben
\uv \leq v^* \leq \bv,
\enn
for $t\in [0,T]$.
The controls on $v^*_T$ imply for $\bw, \uw$
\ben
\bar{w}(T)+\lambda B \leq -\exp(-\frac{x+\lambda C}{\lambda\epsilon_1}) - \exp(-\frac{x+\lambda C}{\lambda\epsilon_2}),\\
\bar{w}(T) +\lambda B\geq -\exp(-\frac{x-\lambda C}{\lambda\epsilon_1}) - \exp(-\frac{x-\lambda C}{\lambda\epsilon_2}),
\enn
hence solving for time $t\leq T$,
\ben
\bar{w}(t)+\lambda B \leq -\exp(-\frac{x+\lambda C}{\lambda\epsilon_1} +\bar{D}_1 (T-t) - \exp(-\frac{x+\lambda C}{\lambda\epsilon_2}+\bar{D}_2(T-t)),\\
\bar{w}(t) +\lambda B\geq -\exp(-\frac{x-\lambda C}{\lambda\epsilon_1}+\bar{D}_1 (T-t)) - \exp(-\frac{x-\lambda C}{\lambda\epsilon_2}+\bar{D}_2(T-t)),
\enn
with $\bar{D}_1, \bar{D}_2$ that depend also on $\bar{\sigma}, \underline{\sigma}$.
A similar estimate hold for $\underline{w}$ with other constants $\underline{D}_1,  \underline{D}_2$.
These controls translate back to $\bar{v}$ and $\underline{v}$, and by comparison to $v$ to yield the result of the Lemma. 

$\hfill \Box$

By convexity of $v^*$, Lemma \ref{finebouds} yields then the following  control on $\dy v^*$:
\begin{lemme}\label{gradv*control}
There exists $\theta_1, \theta_2>0$ such that
\ben
\frac{\theta_1}{-(y+\lambda B)} \leq \dy v^* \leq \frac{\theta_2}{-(y+\lambda B)}.
\enn
Moreover if $\theta=\frac{\theta_2}{\theta_1}>1$ then $\forall \alpha >0, t\leq T, y\in (-\infty, -\lambda B)$, 
\beq\label{increasedyua}
\frac{1}{\dy v^*(-\lambda B-\alpha \theta)}+\alpha \leq  \frac{1}{\dy v^*(-\lambda B-\alpha)}.
\enq
\end{lemme}

{\it Proof.} This is a direct consequence of the previous Lemma \ref{finebouds}. The  accurate control on the upper and lower barrier (same logarithmic growth for upper and lower barrier) yields a  precise estimate of the gradient.

$\hfill \Box$

{\it Proof of Theorem \ref{ubvbm}.}

We can invoke an appropriate result of regularity (see Lieberman \cite{Lieberman}, Lemma 12.13) that state that locally $\dy v^*$ is H\"older continuous in space. Adapted to our case here is the result:

\begin{lemme}[\cite{Lieberman}]\label{Lieberman}
Let $v^*$ solve on $(a, b)\times [0,T]$    equation (\ref{eqonv*}), such that for some $\nu>0$,  $\nu \leq \dy v^* \leq \nu^{-1}$. Then for $\alpha\in (0,1)$, for $t\in [0,T-\tau]$,  for $\omega \subset\subset (a,b)$
\ben
\|\dy v^*\|_{C^\alpha(\omega)} \leq C(\alpha, \nu,\tau,\sigma, b-a,\omega).
\enn
The dependence with respect to $\sigma$ is controlled by $\underline{\sigma}, \bar{\sigma}$,the lower and upper bounds on $\sigma$ and $\sup_{[0,T]\times(a,b)} |\ds\sigma(t,s)| + |\dt\sigma(t,s)|.$
If moreover $v^*(t=T) \in C^{1+\alpha}_{loc}(a,b)$ then $v^* \in C^{1+\alpha}_{loc}([0,T]\times (a,b))$.
\end{lemme}

For $y+\lambda B \in [-\theta, -1]$, consider for some $\alpha>0$ $$u^\alpha(t,y) = v^*(t,-\lambda B + \alpha (y+\lambda B)).$$
Then $\dy u^\alpha(t,y)=\alpha \dy v(t,-\lambda B + \alpha (y+\lambda B))$,
and $u^\alpha$ solves
\beq\label{paraboalpha}
\dt u^\alpha + \frac{\sigma^2(t, \alpha^{-1} \dy u^\alpha) }{2} \frac{\dyy u^\alpha}{(\dy u^\alpha)^2}=\frac{\sigma^2(t, \alpha^{-1} \dy u^\alpha) }{2}.
\enq

We now use Lemma \ref{gradv*control}, which yields that for some positive constants $\theta'_1, \theta'_2$, one has 
$$y+\lambda B\in [-\theta,-1] \Rightarrow \theta'_1\leq \dy u^\alpha \leq \theta'_2 .$$  

Hence, Lemma \ref{Lieberman} applies, and granted we have a bound on $s\ds \sigma(t,s)$ (as assumed in the Theorem), $\dy u^\alpha$ is bounded and H\"older continuous uniformly with respect to $\alpha$. Then using Schauder estimates  (\cite{Lieberman}, see also \cite{WangSchauder} for a very synthetic proof) this gives a uniform control on $\dyy u^\alpha$ and $\dt u^\alpha$  in $C^{\beta}([0,T-\tau]\times [-\theta-\lambda B,-1-\lambda B])$. 
This in turn guarantees that $\dt v^*$ is also uniformly bounded in $[0,T-\tau]\times \R_+^*$, and and since $\dt u(t,s) = -\dt v^*(t,y)$, one obtains the result that $\frac{1}{1-\lambda s^2\dss u}$ is uniformly bounded on $[0,T-\tau]\times \R_+^*$.

We obtain a bound on $s^2\dss u$ from below as a consequence of  Harnack inequality. 
Using the choice of $\theta$ and (\ref{increasedyua}), 
$u^\alpha$ will satisfy uniformly with respect to $\alpha$
\ben
\frac{1}{\dy u^\alpha}(-\lambda B-\theta)+1 \leq  \frac{1}{\dy u^\alpha}(-\lambda B-1),
\enn
and given the  bound on $\dy u^\alpha$, this implies
\beq\label{ualpha}
\dy u^\alpha(-\lambda B-1) \geq \dy u^\alpha(-\lambda B-\theta)+\varepsilon_0,
\enq
for some $\varepsilon_0>0$.
Then, differentiating twice the equation (\ref{paraboalpha}), we obtain that $z=\dyy u^\alpha$ solves an equation of the form
\beq\label{eqw}
\dt z + \dy(A(t,\dy u^\alpha) \dy z) + \dy (C(t,\dy u^\alpha, z) z) = 0,
\enq 
where $A=\frac{\sigma^2(t,\alpha^{-1}\dy u^\alpha)}{2(\dy u^\alpha)^2}$. 
Under the assumption that $s\ds\sigma$ is bounded, the coefficient $A,C$ are uniformly bounded with respect to $\alpha$, and $A$ is also bounded away from 0.
We already now that $z$ is positive, bounded and continuous, moreover by (\ref{ualpha}) we have that 
\ben
\sup\{\dyy u^\alpha(t,y), y\in  [-\lambda B-\theta, -\lambda B-1]\} \geq \frac{\varepsilon_0}{\theta-1}.
\enn
We can now invoke Harnack inequality for the solution of (\ref{eqw}) (see \cite{Lieberman} Theorem 6.27) that implies that, locally,  the supremum of $w$ at a given time $t$ is controlled by the infimum of $w$ at time $t'<t$. Hence the infimum of $w$ has to stay uniformly away from 0.
This in turn implies that $\frac{\dyy v^*}{(\dy v^*)^2}$ remains uniformly bounded away from $0$ on $[0,T-\tau]$, hence that $s^2\dss u$ is bounded away from $-\infty$,
 and we have obtained the following lemma:
\begin{lemme}
For $\tau>0$, $s^2 \dss u$ belongs to $C^{\alpha}_{x}([0,T-\tau]\times\R_+^*)$, and
there exists $\varepsilon_1>0$ depending on $\tau$ such that for $t\in [0,T-\tau]$, $s>0$ 
\ben
-\varepsilon_1^{-1}\leq s^2\dss u(t,s)\leq \Lambda(1-\varepsilon_1).
\enn
\end{lemme}

 We now differentiate the equation with respect to $y$ to obtain $C^3$ regularity for $u^\alpha$. Since $s^2\dss \sigma$ is bounded, we also have that $y \to (\sigma(t,\alpha^{-1} \dy u^\alpha))$ is bounded in $C^{1+\beta}_s(K)$ for $K$ compactly supported in $[0,T)\times (-\infty, -\lambda B)$. We also assume that $s\partial_{st}\sigma$ bounded, hence $u^\alpha$ belongs to $C^{3+\beta}_s([0,T-\tau]\times[-\lambda B-\theta, -\lambda B-1])$. Then we observe that 
\ben
v(t,s)= (u^\alpha(t))^*(s/\alpha),
\enn
hence
\ben
\alpha^2\dss v(\alpha s) &=& \frac{1}{\dyy u^\alpha(\alpha\ds v(\alpha s))},\\
\alpha^3\partial^3_s v(\alpha s) &=& \frac{-\partial^3_y u^\alpha \alpha^2 \dss v(\alpha s)}{(\dyy u^\alpha(\alpha\ds v(\alpha s)))^2}.
\enn
This implies that $s^2\dss v$ and $s^3\partial^3_{s} v$ are bounded, and the same holds then for $u$, given that $\dyy u^\alpha$ is bounded away from $0$ as we just showed.

To prove regularity up to the initial boundary when the initial data is such that $u^\alpha(T)$ is uniformly bounded in $C^{2+\alpha}$ (which is the case if $s^3\partial^3_s\Phi$ is bounded) we just apply standard parabolic regularity (see \cite{Lieberman}, Theorem 5.14).

\subsection{Local interior bounds}

Under minimal assumptions on $\Phi$, one can still establish local interior regularity. The proof of this result is deferred to the Appendix \ref{prooftheolocvol}, as it is a similar to the proof of Theorem \ref{ubvbm}.
\begin{theo} \label{theolocvol} Asssume that $\sigma \in C^2([0,T]\times \R_+^*) $ with $\underline{\sigma}\leq \sigma \leq \bar{\sigma}$ for $(\underline{\sigma}, \bar{\sigma})$ positive constants.
Let the terminal payoff $\Phi$ satisfy  
$\lambda s^2 \dss \Phi \leq 1$ and condition (\ref{kernel}) with $\sigma = \bar{\sigma}$.
Then any classical solution $u$ to (\ref{main-lambda-dep1},\ref{main-lambda-dep2},\ref{main-lambda-dep3}) satisfies:
For all compact set $K\subset [0,T)\times \R_+^*$, $\alpha \in(0,1)$
\ben
&&\|u\|_{C^{3+\alpha,\frac{3}{2}+\alpha/2}_{s,t}(K)} \leq Q,\\
&&-Q \leq  s^2\dss u \leq 1/\lambda -1/Q,\\
&&Q=Q(K,\Phi,\underline{\sigma}, \bar{\sigma}, \|\sigma\|_{C^{1+\alpha,\demi(1+\alpha)}_{s,t}(K)},\alpha).
\enn
If $\sigma$ is only $C^{\alpha, \alpha/2}_{t,s,loc}([0,T]\times \R_+^*)$ then $u$ is locally bounded in $C^{2+\alpha, 1+\alpha/2}_{s,t,loc}$ with similar bounds.

\end{theo}
{\it Remark.} 
The bound $Q$ can be be made dependent only on $\underline{\Phi}, \bar{\Phi}$ for $\Phi$ varying in $\{\Phi: \lambda s^2 \dss\Phi \leq 1, \underline{\Phi}\leq \Phi \leq \bar{\Phi}\}$.

In view of Theorem \ref{theocstvol}, the condition on $\Phi$ is sharp, as if $\sigma \equiv \bar\sigma$ and (\ref{kernel}) does not hold, the solution will blow up form some time $t_0>0$.

\subsection{Initial regularity}
We also mention that regularity holds up to the initial time on cylinder "above" areas where $u$ is smooth. This result will be useful to define the stock's dynamic as a proper martingale up to time $T$ (see Theorem \ref{local-rep-2}).
\begin{theo}\label{cylinder}
In addition to the assumptions of Theorem \ref{theo-main}, if $\Phi \in C^{2+\alpha}([a,b])$ for some interval $[a,b]$ where it satisfies  (\ref{contrainte-strict}), then 
\beq\label{eqcylinder} 
u \in C^{2+\alpha, 1+\alpha/2}_{s,t,loc}([0,T]\times (a,b)).
\enq
If moreover $\Phi\in C^{3+\alpha}([a,b])$, then $u\in C^{3+\alpha}((a,b)\times [0,T])$.
\end{theo} 

{\it Proof.} 
It uses a classical cutoff argument. Multiplying $v^*$ by a cutoff function $\eta(y)$ compactly supported in $(a,b)$, $h=v^*\eta$ solves 
\ben
\dt h + A \dyy h +B = A(\dy v^* \dy \eta + v^*\dyy \eta),
\enn
for $A=A(t,\dy v^*), B=B(t,\dy v^*)$. One already has from Lemma \ref{Lieberman} a global $C^{1+\alpha}_{loc}([0,T] \times (a,b))$ bound on $v^*$, from Lemma \ref{Lieberman}. Since $h$ is $C^{2,\alpha}$ smooth on the parabolic boundary of $[a,b]\times[0,T]$, classical Schauder regularity (see again Lieberman \cite{Lieberman}, Theorem 5.14) applies up to the boundary and yields the desired result.

$\hfill \Box$

\subsection{Construction of solutions}

We now prove the existence of solutions, using the a priori bounds and the continuity method:
\begin{prop}\label{exiphi}
Let $\Phi, \sigma$ satisfy the assumptions i) to v) of Theorem \ref{ubvbm}. 
Then there exists a classical solution to (\ref{main-lambda-dep1}, \ref{main-lambda-dep2}, \ref{main-lambda-dep3}).
\end{prop}

{\it Proof.} We consider for $\varepsilon \in [0,1]$  $\sigma\ep$ defined as follows:
\ben
\sigma\ep&=&1 \text{ if } t\geq T(1-\varepsilon),\\
&=&((1-\varepsilon)\sigma(t,s) + \varepsilon) \text{ if } t\leq T (1-2\varepsilon),\\
&=& 1 +  \varphi\ep(t)(1-\varepsilon)(\sigma(t,s)-1)\text{ if } t\in [T(1-2\varepsilon), T(1-\varepsilon)],
\enn
where $\varphi\ep\in C^\infty$ is non-negative, non-increasing and satisfies:
\ben
\varphi\ep&=&0 \text{ if } t\geq T(1-\varepsilon),\\
&=& 1 \text{ if } t\leq T(1-2\varepsilon).
\enn
We start from $\varepsilon=1$, and consider the derivative of $u\ep$ solution to (\ref{main-lambda-dep1}) with respect to $\varepsilon$:
$u\epu=\frac{d}{d\varepsilon}u\ep$ solves
\ben
\dt u\epu + \demi (\sigma\ep)^2\epu\frac{\gamma\ep}{1-\lambda\gamma\ep} + \demi  (\sigma\ep)^2
\frac{s^2\dss u\epu}{(1-\lambda \gamma\ep)^2}=0,
\enn
with $u\ep(T)=0$,
where $\gamma\ep=s^2\dss u\ep$ and $(\sigma\ep)^2\epu=\frac{d}{d\varepsilon}(\sigma\ep)^2.$
If for some $\nu'(\varepsilon)$ there holds
\beq\label{sec5nu}
&&\nu' \leq 1-\lambda\gamma\ep \leq 1/\nu',\\
&&\gamma\ep\text{ is globally Lipschitz with respect to }s,\label{sec5nu2}
\enq
then one can define $S^{\epsilon , t,s}$ the unique strong solution on $[t,T]$ to  
\ben
d S^{\epsilon , t,s}_{t'} &=& S^{\epsilon , t,s}_{t'}\frac{\sigma\ep}
{1-\lambda\gamma\ep}dW^{\bP}_{t'},\\
S^{\epsilon, t,s}_{t}&=&s,
\enn
and $u\epu$ can be found by the following representation formula
\ben
u\epu(t,s) = \E^\bP\left(\int_t^T \demi (\sigma\ep)^2\epu\frac{\gamma\ep}{1-\lambda\gamma\ep}(t,S^{\epsilon , t,s}_{t'}) dt'\right),
\enn
and $u\epu\in C^{1+\beta, 3+\beta}_{t,s,loc}$.
Hence the linearized operator is invertible at a point $u\ep$ satisfying (\ref{sec5nu}, \ref{sec5nu2}), and from the implicit functions theorem one can find a solution $u^{\varepsilon_2}\in C^{1+\alpha, 3+\alpha}_{t,s,loc}$ for $\varepsilon_2$ close to $\varepsilon$. 

Then, under the assumption of Theorem \ref{ubvbm}, (\ref{sec5nu}, \ref{sec5nu2}) hold on $[0,T(1-\varepsilon)]$, while, as $\sigma\ep\equiv 1$ on  $[T(1-\varepsilon), T]$,  Theorem \ref{theocstvol}, yields that (\ref{sec5nu},\ref{sec5nu2}) hold on $[T(1-\varepsilon), T]$.  Therefore one can apply the continuity method (see \cite{GT})  to build the curve $u\ep, \varepsilon \in [1,0]$, and $u\ep$ enjoys uniformly  the a priori estimates of Theorem \ref{ubvbm}. Finally $u^0$ solves (\ref{main-lambda-dep1}, \ref{main-lambda-dep2}, \ref{main-lambda-dep3}).

$\hfill \Box$

\subsection{Existence and uniqueness results}
We conclude with the main existence, regularity result:
\begin{theo}[Existence and regularity]\label{theo-main}
\begin{enumerate}
\item Let $\Phi, \sigma$ satisfy the assumptions i) to iv) of Theorem \ref{ubvbm}. Then there exists a unique classical solution to (\ref{main-lambda-dep1}, \ref{main-lambda-dep2}, \ref{main-lambda-dep3}). 

\item If $\Phi, \sigma$ satisfies only the assumptions of Theorem \ref{theolocvol}, then there exists a classical solution to (\ref{main-lambda-dep1}, \ref{main-lambda-dep2}, \ref{main-lambda-dep3}). 

\end{enumerate}
\end{theo}
{\it Remarks.} Those solutions will then naturally enjoy the a priori estimates of Theorem \ref{ubvbm} or \ref{theolocvol}, as they are satisfied for any classical solution.

Again, in view of Tychonov's counterexamples \cite{Tycho}, such a uniqueness result in $\R_+^*$  seems surprising without any growth condition, but recall that in our definition of classical solutions (Definition \ref{def-classical}) we embed an assumption of semi-concavity for $u$.

{\it Proof.} One can approximate $\Phi$ by a sequence $\Phi^\nu$ satisfying the assumptions i) to v)of Theorem (\ref{ubvbm}). By letting $\nu$ go to 0 the interior estimates do not depend on $\nu$, and using a standard compactness argument the sequence of solutions $u^\nu$ converges locally uniformly, and the a priori estimates pass to the limit. 

We prove now the uniqueness part, and for this we need the following result whose proof is deferred to the appendix \ref{prooftheo-comp}.
\begin{theo}\label{theo-comp}
Let $u_1, u_2\in C_{s,t}^{3,1}([0,T]\times \R_+^*)$ satisfy, for some $C>0$,
\ben
&&-C\leq s^2\dss u \leq \lambda^{-1} - 1/C,\\
&& |s^2\dss u| + |s^3\partial_{sss}u| +|s\ds \sigma| \leq C.
\enn
Assume then that
\ben
\dt u_1 + \frac{\sigma^2(t,s)}{2}F(s^2\dss u_1) &\leq& 0,\\
\dt u_2 + \frac{\sigma^2(t,s)}{2}F(s^2\dss u_2) &\geq& 0,
\enn
with $F$ as in (\ref{main2}). 
Then if $u_1(T) \geq u_2(T)$, $u_1(t) \geq u_2(t)$ on $[0,T]$.
\end{theo}

We already know that the a priori estimate 
of Theorem \ref{ubvbm} holds. 
Then by the comparison result of Theorem \ref{theo-comp} applied on $[0,T-\tau]$ (which is allowed using the interior {\it a-priori} estimates), one obtains that
\ben
t\to \sup_{s>0} |u_1(T-t,s)-u_2(T-t,s)|
\enn
is non-increasing. 
We can conclude if for the two solutions there holds 
\beq\label{lim}
\lim_{t\to T} \|u(t,\cdot)-\Phi(\cdot)\|_{L^\infty(\R_+^*)} = 0,
\enq
This comes by considering the rescaled solution $u^\alpha$ introduced in the proof of Theorem \ref{ubvbm}. By a compactness argument, $u^\alpha(t,)$ converges to $u^\alpha(T,\cdot)$ in $C^0([-\lambda B - \theta, -\lambda B - 1])$ as $t\to 0$, uniformly with respect to $\alpha$, which implies (\ref{lim}).

$\hfill \Box$

\section{Representation of the solution}

We consider as above $W^\bP_t$ a standard Brownian motion on a filtered probability space $(\Omega, \bF, {\cal F}_t, \bP)$, and $S$ to be the solution to (\ref{dSver1}, \ref{dSver2}, \ref{dSver3}), i.e.
\ben\label{dS}
\frac{dS_t}{S_t}=\frac{\sigma(t,S_t)}{1-\lambda\gamma(t,S_t)} dW^\bP_t,
\enn
with $S_0$ as initial condition.
Following the notations introduced previously,  we introduce $V(t,s) = v^*(t,\ds v)$ with $v(t,s)=-\ln(s)-u -1$, and, and  $\cS$ is the inverse of $V$ with respect to $s$: 
\beq\label{defV}
V(t,s)&=&\ln(s)+\lambda(u(t,s)-s\ds u(t,s)),\\
\cS(t,\cdot)&=&[V(t,\cdot)]^{-1},\label{defV2}
\enq
and (\ref{defV2}) has to be understood in the sense of  Proposition \ref{definverse}.

\subsection{A maximum principle for the second derivative}
We state this result of independent interest for $F$ as in (\ref{main-lambda-dep2}), but it is valid for a large range of non-linear diffusions. The result will be used in the proof of Theorem \ref{local-rep-2}.
\begin{theo}\label{max-princ-2nd}
Let $\sigma\in C^{1,1}_{t,s}([0,T]\times \R_+^*)$ satisfy
 $\underline{\sigma}\leq \sigma \leq\bar{\sigma}$ for some positive constants $\underline{\sigma},\bar{\sigma}$,  with  $\sup_{s>0,t\geq 0}\{|\dt \sigma(t,s)|\}<+\infty$ .
Let $u$ be a classical solution to (\ref{main-lambda-dep1}, \ref{main-lambda-dep2}, \ref{main-lambda-dep3}), and assume that $F(\gamma(t,s))$ is uniformly bounded by above.
Then for all $t\in [0,T], s>0$, letting $S^{t,s}$ be the solution of (\ref{dSver1}, \ref{dSver2}, \ref{dSver3})
\beq\label{FKV}
&(\sigma^2F(s^2\dss u))(t,s)&=\\
&&\nonumber\E^\bP\left\{(\sigma^2F(s^2\dss u))(T,S^{t,s}_T) e^{-\int_t^T\frac{\dt(\sigma^2)}{\sigma^2}(t',S^{t,s}_{t'})}dt'\right\}.
\enq
If one does not assume that $F$ is globally bounded, there holds  
\beq\label{FKVineq}
&(\sigma^2F(s^2\dss u))(t,s)&\geq\\
&&\nonumber\E^\bP\left\{(\sigma^2F(s^2\dss u))(T,S^{t,s}_T) e^{-\int_t^T\frac{\dt(\sigma^2)}{\sigma^2}(t',S^{t,s}_{t'})}dt'\right\}.
\enq

\end{theo}

{\it Proof.} 
We proceed by approximation. On $Q^\nu=[0,T-\nu]\times [\nu,1/\nu]$ one can choose $\sigma^\nu$ $C^\infty$ smooth and close to $\sigma$, so that the solution $u^\nu$ to (\ref{main-lambda-dep1}) with $u^\nu=u$ on the parabolic boundary of $Q^{\nu}$
is $C^\infty$ smooth in the interior of $Q^\nu$ and $C^{2+\alpha, 1+\alpha/2}_{t,s}$ globally on $Q^\nu$. Then $u^\nu$  is smooth enough to 
differentiate twice the equation, and $V^\nu=(\sigma^\nu)^2(t,s)F(\gamma^\nu)$ solves
\beq
\dt V^\nu + \frac{(\sigma^\nu)^2s^2}{2}F'(\gamma^\nu) \dss  V^\nu = \frac{\dt ((\sigma^\nu)^2)}{(\sigma^\nu)^2}V^\nu\label{eqongamma},
\enq
which, by Schauder regularity implies $V^\nu\in C^{2,\alpha, 1+\alpha/2}_{s,t,loc}$, uniformly and since $\sigma^\nu \in C^{1,1}$ implies that $F(\gamma^\nu)$ and hence $\gamma^\nu$ are (locally in $t,s$, uniformly with respect to $\nu$) Lipschitz. 
As we are on a bounded domain, by uniqueness $u^\nu$ converges to $u$, and $V^\nu$ to $V$. The stochastic differential equation (\ref{dSver1},\ref{dSver2},\ref{dSver3}) admits then a unique local strong solution, and $V$ is uniquely defined by the representation formula:
\ben
V(t,s)=\E^\bP \left\{V(T,S^{t,s}_{\tau_\nu}) e^{-\int_t^{\tau_\nu}\frac{\dt(\sigma^2)}{\sigma^2}}\right\},
\enn
where $\tau_{\nu}$ is the first exit time outside of $Q_\nu$.
We now let $\nu$ go to 0, as $\tau_{\nu} \to T$ $\bP -$ a.s.,
assuming an a priori  bound on $\frac{\dt(\sigma^2)}{\sigma^2}, V$, the equality above remains true by uniform integrability. 
If the upper bound on $V$ is not assumed, there still remains the lower bound as $F\geq -1/\lambda$, and by Fatou's lemma, the inequality (\ref{FKVineq}) holds.

$\hfill \Box$

We now introduce a modified process $Y_t$ that will be key to study the properties of the process $S_t$.
\begin{prop}\label{Backlund Y}
Let $\Phi, \sigma$ satisfy the assumptions of Theorem \ref{theolocvol}. Let $u$ be a $C^{3,1}_{s,t,loc}$ classical solution to (\ref{main-lambda-dep1},\ref{main-lambda-dep2},\ref{main-lambda-dep3}), and $S_t$ be a local strong solution to (\ref{dSver1}).  Let $V(t,s)$ be defined from $u$ as above.  Consider the process 
\beq\label{defY} 
Y_t &=& V(t,S_t)\\
&=& \ln(S_t) + \lambda(u(t,S_t) - S_t\partial_s u(t,S_t)).\nonumber
\enq
Then $Y$ satisfies
\beq
dY_t=\sigma\left((dW^\bP_t - s\ds \sigma dt) +\frac{s\ds\sigma -\sigma}{1-\lambda\gamma} dt\right)+\frac{\sigma^2}{2}dt.\label{dY}
\enq
\end{prop}

{\it Remark.} When $\lambda =0$ we recover that $d\ln(S_t)=\sigma dW^\bP_t - \sigma^2/2 dt$.

{\it Proof.} We have that
\ben
du(t,S_t) &=& \ds u(t,S_t) dS_t + \frac{\lambda \gamma^2 \sigma^2}{2(1-\lambda\gamma)^2}dt,\\
d(S_t\ds u(t,S_t))&=&(\ds u +S_t\dss u)dS_t + (\dss u  + \demi S_t \partial_{sss}u)\frac{\sigma^2S_t^2}{(1-\lambda\gamma)^2} + S_t\partial_{ts}u,\\
d\ln(S_t)& =& \frac{dS_t}{S_t} - \demi\frac{\sigma^2}{(1-\lambda\gamma)^2},\\
-\partial_{ts}u&=& \demi\ds(\sigma^2)\frac{\gamma}{1-\lambda\gamma}+ \frac{\sigma^2}{2}\frac{2S_t\dss u + S_t^2\partial_{sss}u}{(1-\lambda\gamma)^2}. 
\enn
Rearranging the terms we obtain that
\ben
\frac{1}{\lambda} dY_t &=& (\frac{1}{\lambda}-\gamma)\frac{dS}{S}
 -\frac{\sigma^2}{\lambda(1-\lambda\gamma)}dt+ \demi\frac{\gamma s \ds (\sigma^2)}{1-\lambda \gamma})dt,\\
 &=&\frac{\sigma}{\lambda}dW^\bP_t -\frac{\sigma^2}{\lambda(1-\lambda\gamma)}dt + \frac{\sigma s\ds\sigma}{\lambda}(-1+\frac{1}{1-\lambda\gamma})+\frac{\sigma^2}{2\lambda}dt,
\enn
and the result follows.

$\hfill \Box$

\subsection{Dual representation formula and martingale property under $\bP$}
Based on the variational formulation (\ref{variat}), we now have a representation result for the solution $u$, as well as a martingale property result for $S_t$ solution of (\ref{dSver1}, \ref{dSver2}, \ref{dSver3}).

\begin{theo}\label{local-rep-2}
Let  $\cS_T$  be as in (\ref{defcS}).
Assume that (\ref{contrainte-large}), (\ref{kernel}) hold and that
\beq\label{finite0}
\cS_T(y)\text{ is finite for every }y\in \R,
\enq
that $\underline{\sigma} \leq \sigma(t,s) \leq \bar{\sigma}$ for some $\usig, \bsig$ positive, and that $(s+s^2)|\ds\sigma|$ is bounded.
Let $u$ be a classical solution to (\ref{main-lambda-dep1}, \ref{main-lambda-dep2}, \ref{main-lambda-dep3}).
Then,
\begin{itemize}
\item[-] There exists a unique classical solution $S_t$ to (\ref{dSver1}, \ref{dSver2}, \ref{dSver3}) defined on $[0,T]$, 
\item[-] $S_t$ is a martingale on $[0,T]$,
\item[-] (\ref{dualverif}) holds
\beq\label{dualveriftheo}
u(0,S_0) = \E^\bP\left(\Phi(S_{T})-\frac{1}{2\lambda}\int_0^{T}\left(\frac{\sigma(t,S_t)}{1-\lambda\gamma(t,S_t)} - \sigma(t,S_t)\right )^2dt\right),
\enq
(although $\E^\bP(\Phi(S_T)$ may not be finite, and then (\ref{dualveriftheo}) should be understood in the sense of (\ref{repgene})), and moreover
\beq\label{supa}
u(0,S_0) = \sup_{a\in {\cal A}_T}\E^\bP\left(\Phi(S^a_{T})-\frac{1}{2\lambda}\int_0^{T}((a_t)^{1/2} - \sigma(t,S_t) )^2dt\right),
\enq
where $a, {\cal A}_T$ and $S^a$ are defined in (\ref{defA}, \ref{defsa1}, \ref{defsa2}),
\item[-]
\beq\label{vide}
&&\bP\left\{\lambda S_T^2\dss\Phi(S_T)=1\right\}=0,\\
&&\E^\bP\left(\frac{1}{1-\lambda S_T^2\partial_{ss}\Phi(S_T)}\right)<+\infty.
\enq
\item[-] Identity (\ref{dualverif}) holds for any $\Psi\leq \Phi$ such that 
$(\Psi-\Phi)(1-\lambda s^2\dss \Phi)=0$, 
\item[-] If $\Phi(s)\leq C(1+s) + (\varepsilon_0 -\lambda)\ln(s)$ for some $C,\varepsilon_0>0$ then 
\beq\label{finitevarianceloc}
\E^\bP\left(\int_0^T \frac{\sigma^2(t,S_t)}{(1-\lambda\gamma(t,S_t))^2}dt\right) < + \infty,
\enq
and $\E^\bP(\Phi(S_T))$ is finite.
\end{itemize}
\end{theo}
{\it Remark.}
As we will see, condition (\ref{finite0})  implies $\lim_{s\to+\infty} V_T(s)=+\infty$ and thus implies to be in the second case of point 4 in Proposition \ref{definverse}.

{\it Proof.} 
%
Consider the stopping time 
\beq\label{deftaunu}
\tau_\nu=\inf\{t\in [0,T), -\frac{1}{\lambda}+\nu \leq F(\gamma(t,S_t)) \leq \nu^{-1}\}\wedge T-\nu,
\enq
and define the stopped process $S_t^\nu=S_{t\wedge\tau_\nu}$ the  process stopped at time $\tau_\nu$. We then have the following Proposition:
\begin{prop}\label{Girsanovloc} 

The process $S_t^\nu$ defined above is a martingale on $[0,T]$, and 
under the probability $\bQ^\nu$ given by $$\frac{d\bQ^\nu}{d\bP}|_{\cal F _t}=\frac{S_t^\nu}{S_0},$$ $Y_t$ satisfies on $[0,
\tau_\nu]$
\beq\label{dYgirs}
dY_t = \sigma dW^{\bQ^\nu} +\left(s\sigma\ds\sigma (-1+\frac{1}{1-\lambda \gamma}) +   \frac{\sigma^2}{2}\right)dt , 
\enq
for $W^{\bQ^\nu}$ a Brownian motion under ${\bQ^\nu}$. If moreover $S$ is a martingale up to time $T$, then we define $\bQ$  accordingly and (\ref{dYgirs}) holds up to time $T$.
\end{prop}

{\it Proof.} The proof is a straightforward application of Girsanov's Theorem (see \cite{KaratzasShreve}).

$\hfill \Box$

To prove (\ref{vide}), we use the representation formula (\ref{FKV}). Given the uniform bounds we have on $u$ and $F(\gamma)$ from Theorem \ref{theolocvol}, we claim that there exists a  constant $C$ independent of $\varepsilon$ such that 
\beq\label{borneP}
\bP\left\{\sup_{\tau \in [0,T-\varepsilon]} \{F(\gamma(t,S_t))\} \geq M\right\} \leq \frac{C}{M}.
\enq
Indeed, if this is not true, considering the stopping time $\tau_\nu$,
 by formula (\ref{FKV}) one can show that $F(\gamma(t=0,S_0)) \geq C$ for any $C>0$, which contradicts the regularity result of Theorem \ref{theolocvol}. Then almost surely, $\limsup_{t\to T^-} F(\gamma(t,S_t))$ is finite, which shows that almost surely $S_T^2\dss\Phi(S_T) < 1$, and moreover that almost surely $S_t$ is continuous up to time $T$.

To prove that $S_t$ is a martingale up to time $T$, we consider the stopped process $S^\nu_t$. By the result of Theorem (\ref{theolocvol}), $\tau_\nu$ goes to $T$ $\bP$ a.s. as $\nu$ goes to 0. The sequence $S_t^\nu$ is a sequence of martingales up to time $T$ that satisfies $\E^\bP(S_t^\nu)=S_0$ for all $\nu>0, t\in [0,T]$. If we can show that the family $S_t^\nu$ is equi-integrable, and that it converges to $S_t$, then $S_t$  satisfies $\E^\bP(S_t)=S_0$, and by standard arguments, as $S_t$ is non-negative, this implies that $S_t$ is a martingale up to time $T$. 
We consider the family  $V^\nu(t,s)=V(t\wedge\tau_\nu,s)$, and its inverse $\cS^\nu(t,y)= \cS(t\wedge\tau_\nu,y)$, and we have using Proposition \ref{Girsanovloc}
\ben
\E^\bP(S_t^{\nu}\1_{S_t^\nu\geq M})&=& S_0\E^{\bQ^\nu} (\1_{S_t^\nu\geq M})\\
&=& S_0\bQ^\nu \{\cS^\nu(t,Y_0+Z^\nu_t+A^\nu_t  ) \geq M\}\\
&=& S_0\bQ^\nu \{Y_0+Z^\nu_t+A^\nu_t \geq V^\nu(t,M)\},
\enn
where 
\ben
Z^\nu_t  &=& \int_0^{t\wedge\tau_\nu} \sigma(t,S_t^\nu)dW^{\bQ^\nu}_t,\\
A^\nu_t &=& \int_0^{t\wedge\tau_\nu} \left( s\sigma\ds\sigma (-1+\frac{1}{1-\lambda \gamma}) +   \frac{\sigma^2}{2}\right)dt.
\enn

First we observe that 
by Chebyshev's inequality $\E^\bP(S_t^{\nu}\1_{S_t^\nu\geq M}))$ can be bounded by 
$(V^\nu(t,M))^{-1} \E^{\bQ^\nu}(Y_0+|Z^\nu_t|+|A^\nu_t|)$.
Then we have
\begin{lemme}
Under assumptions (\ref{finite0}), (\ref{kernel}),  
$V^\infty(s) = \inf\{V(t,s), t \in [0,T]\}$ satisfies $\lim_{s\to \infty}V^\infty=+\infty$. 
\end{lemme}
{\it Proof.} Indeed remember that $\cS$ is the inverse of $V$ and that $1/\cS=\dx w$. This is then an  consequence of Lemma \ref{lemmedur} combined with the concavity of $w$: under assumption (\ref{finite0}), $\dx w$ is bounded away from $0$ on every  set  $[0,T]\times A$, $A$ bounded

$\hfill \Box$

Finally  $Z^\nu$ is   bounded in $L^1(\bQ^\nu)$ uniformly with respect to $t,\nu$ and if we have a uniform $L^1$ bound on $A^\nu$, we have a uniform bound on $\E^\bP(S_t^{\nu}\1_{S_t^\nu\geq M}), t\in [0,T], \nu >0$ which shows the equi-integrability of the family. Having observed above that $S_t$ is almost surely continuous up to time $T$, $S_T^\nu$ converges a.s. to $S_T$, hence $E^\bP(S_T)=S_0$ which allows to conclude.
It thus remains to show that
\begin{lemme}\label{sigl1}
Under the assumptions of Theorem \ref{local-rep-2}, $A^\nu_t$ is  bounded in $L^1(\bQ^\nu)$, uniformly with respect to $t,\nu$.
\end{lemme}
{\it Proof.} We will use Theorem \ref{max-princ-2nd} and the inequality (\ref{FKVineq}).
We need to bound 
\ben
&&\E^{\bQ^\nu}\left(\int_0^{t\wedge\tau_\nu} S_{t'} \sigma|\ds\sigma|\frac{1}{1-\lambda\gamma}dt'\right)\\
&=& S_0^{-1} \E^\bP\left(\int_0^{t\wedge\tau_\nu} S^2_{t'} \sigma|\ds\sigma|\frac{1}{1-\lambda\gamma}dt'\right),
\enn
which, assuming a bound on $s^2\ds\sigma$ is equivalent to a bound on
$\E^\bP(\int_0^{t\wedge\tau_\nu} F(\gamma(t',S_{t'}))dt$, and which follows directly from the fact that
\ben
\E^\bP\left(\int_0^{t\wedge\tau_\nu} F(\gamma(t',S_{t'})dt'\right)&=&\int_0^{t} \E^\bP(F(\gamma(t',S_{t'}\wedge\tau_{\nu}))-\nu^{-1}\1_{t'\geq \tau_{\nu}}) dt' \\
&&\leq C(\sigma) T F(\gamma(0,S_0)),
\enn
from Theorem \ref{max-princ-2nd}.
Since $F(\gamma(0,S_0))$ is bounded  by Theorem \ref{theolocvol}, the result follows. 

$\hfill \Box$

To prove (\ref{finitevarianceloc}) we use again the stopped process $S^\nu_t$.
Then for $\nu>0$, (\ref{dualverif}) holds up to $\tau_\nu$, i.e.
\ben
u(0, S_0) = \E^\bP\left(u(\tau_\nu, S_{\tau_\nu})-\frac{1}{2\lambda}\int_0^{\tau_\nu}( \sigma^\gamma-\sigma)^2dt\right),
\enn
where $\sigma^\gamma(t,s)$ is as in (\ref{dSver2}).
Under our assumptions $u$ satisfies for some $C>0$, $u(t,s)\leq C(1+s) + (\varepsilon_0 - \frac{1}{\lambda})\ln(s)$. This combined with the above identity implies that 
\ben
u(0,S_0) \leq C(1+S_0)&-&\demi\varepsilon_0 \E^\bP\left(\int_0^{\tau_\nu} (\sigma^\gamma)^2dt\right)\\
&+& \frac{1}{2\lambda}\E^\bP\left(\int_0^{\tau_\nu} (\sigma^\gamma)^2 - \left(\sigma^\gamma-\sigma\right)^2dt\right),
\enn
hence, since $u(0,S_0)$ is finite, and using Cauchy-Schwartz's inequality, $\E^\bP\left(\int_0^{\tau_\nu}(\sigma^\gamma)^2dt\right)$
is uniformly bounded as $\nu$ goes to 0. 


We now prove that identity (\ref{dualverif}) holds even when $u$ is not smooth up to time $T$. We consider the stopping times $\tau_\nu$. Then there will hold by It\^o's formula
\ben
u(0,S_0)&=&\E^\bP\left(u(\tau_n,S_{\tau_\nu}) -\frac{1}{2\lambda}\int_0^{\tau_\nu}(\sigma^\gamma-\sigma)^2dt\right)\\
&=&\E^\bP\left(v(\tau_\nu,S_{\tau_\nu})-\frac{1}{2\lambda}\ln(S_{\tau_\nu})-\frac{1}{2\lambda}\int_0^{\tau_\nu}(\sigma^\gamma-\sigma)^2dt\right)\\
&=&\E^\bP\left(v(\tau_\nu,S_{\tau_\nu})+\frac{1}{\lambda}\int_0^{\tau_\nu}\sigma^\gamma\sigma - \frac{1}{2} \sigma^2dt\right),
\enn
where $v(t,s)=\frac{1}{\lambda}\ln(s)+u$.
Arguing as Lemma \ref{sigl1}, $\E^\bP(\int_0^T\sigma^\gamma(t,S_t) dt)$ is bounded under the assumption that $s\ds \sigma$ is bounded. By monotone convergence, $\int_0^{\tau_\nu}\sigma^\gamma\sigma dt$ converges thus to $\int_0^{T} \sigma^\gamma\sigma dt$. Then under (\ref{contrainte-large}), $v$ is  bounded above by $C(1+s)$, hence by Fatou's lemma
\ben
\limsup \E^\bP\left(v(\tau_\nu,S_{\tau_\nu}) -C(1+S_{\tau_\nu})\right)&\leq& \E^\bP(v(T,S_T) -C(1+S_T))\\
&=& \E^\bP(v(T,S_T) -C(1+S_0))
\enn
as we know already that $S$ is a martingale.
We then have the first inequality
\ben
u(0,S_0) \leq \E^\bP\left(v(T,S_{T})+\frac{1}{\lambda}\int_0^{T}\sigma^\gamma\sigma - \demi \sigma^2dt\right).
\enn
On the other hand we know that $v$ is concave hence by Jensen's inequality (and using again that $S$ is a martingale)
\ben
\E^\bP (v(\tau_\nu,S_{\tau_\nu})) &\geq& \E^\bP (v(\tau_\nu,S_T)-\ds v(\tau_\nu, S_{\tau_\nu})(S_T-S_{\tau_\nu}))\\
&=& \E^\bP(v(T,S_T)) + \E^\bP( v(\tau_\nu,S_T) - v(T,S_T) )\\
&=& \E^\bP(v(T,S_T)) + \E^\bP( v(\tau_\nu,S_T) - v(T,S_T) -C(\tau_\nu-T) )  +  C\E^\bP(\tau_\nu-T) 
\enn where $C$ is chosen so that $v-Ct$ is non increasing (such a $C$ exists since $F$ is bounded by below). By monotone convergence the second term goes to 0 and the third converges easily to 0, which shows that  
\ben
u(0,S_0) \geq \E^\bP\left(v(T,S_{T})+\frac{1}{\lambda}\int_0^{T}(\sigma^\gamma\sigma - \demi \sigma)^2dt\right),
\enn
and equality thus follows:
\beq\label{repgene}
u(0,S_0) = \E^\bP\left((\Phi+\frac{1}{\lambda}\ln)(S_{T})+\frac{1}{\lambda}\int_0^{T}(\sigma^\gamma\sigma - \demi \sigma)^2dt\right).
\enq
{\it Remark.} As stated in the Theorem, this equality holds true even if $\E^\bP((\Phi+\frac{1}{\lambda}\ln)(S_{T}))$ is always finite under the assumptions of the Theorem, and then (\ref{repgene}) is equivalent to (\ref{dualveriftheo}).

Following the same lines, for any element $a$ of ${\cal A}_T$ one can reproduce the computations of section \ref{dualsection} and find that
\ben
&&\E^\bP\left(u(\tau_\nu,S_{\tau_\nu}) -\frac{1}{2\lambda}\int_0^{\tau_\nu}(\sigma^\gamma-\sigma)^2dt\right)  \\
&\geq& \E^\bP\left(u(\tau_\nu,S^a_{\tau_\nu}) -\frac{1}{2\lambda}\int_0^{\tau_\nu}(a^{1/2}-\sigma)^2dt\right).
\enn 
It is straigthforward to show that the second part converges to 
\ben
\E^\bP\left(\Phi(S^a_T) -\frac{1}{2\lambda}\int_0^T(a^{1/2}-\sigma)^2dt\right)
\enn
when $\nu$ goes to $0$, and this shows one part of (\ref{supa}). To show the other side (that $u$ indeed realizes the supremum), one considers again the stopped process and $a_{n,t}=\sigma^\gamma\1_{t\leq\tau_\nu}$. Then $S^{a_\nu}_{t}=S_{t\wedge\tau_\nu}$, $a_\nu$ belongs to ${\cal A}_T$, and the same arguments as above show that 
\ben
u(0,S_0)=\lim_{\nu \to 0}\E^\bP\left
(\Phi(S^{a_\nu}_T)-\frac{1}{2\lambda}\int_0^T(a_\nu^{1/2}-\sigma)^2dt\right).
\enn

The statement about identity (\ref{dualverif}) is a direct consequence of (\ref{vide}).

 $\hfill \Box$

\subsection{Black-Scholes representation formula, and explicit formulation of the density}
We go back to the equation (\ref{eqonw}). As has already been observed $w=[v^*]^{-1}$ is given by
\ben
w(t,x) =  -\frac{1}{{\cal S}(t,x)}-\lambda\ds u(t,{\cal S}(t,x)),
\enn
where $\cS$ has been defined in (\ref{defV}), and $w$ satisfies 
\beq
\dt w + \demi\sigma^2(t,\frac{1}{\dx w})(\dxx w +\dx w) & =&0.\label{w1}
\enq
We consider for $\tilde{\bP}$ a probability on $(\Omega, \bF, ({\cal F}_t)_{t\geq 0})$, $W^{\tilde{\bP}}_T$ a $\tilde{\bP}$-Brownian motion and ${\cal Y}_t$ the  solution to 
\beq
d{\cal Y}_t&=&\sigma(dW^{\tilde{\bP}}_t+\sigma/2 dt),\label{defcalY}\\
\sigma&=&\sigma(t,{\cal S}(t,{\cal Y}_t)),\nonumber\\
{\cal Y}_{0} &=& V(0,S_0).\nonumber
\enq
Under the assumptions of Theorem \ref{theolocvol},  $w$ has enough regularity to check that $w(t,{\cal Y}_t)$ is a local martingale on $[0,T)$. We have then the following representation result, which we call a Modified Black-Scholes representation:

\begin{theo}[Local volatility]\label{local-rep}
Let $u$ be a classical solution to (\ref{main-lambda-dep1}, \ref{main-lambda-dep2}, \ref{main-lambda-dep3}), where $\sigma$, $\Phi$ satisfy the assumptions of Theorem \ref{theolocvol}.  Let $w$ be defined from $u$ as above.
Let ${\cal Y}_t$ be defined as in (\ref{defcalY}). Then $w(t,{\cal Y}_t)$ is a martingale up to time $T$, 
\beq
-\frac{1}{S_0}-\lambda \ds u(t_0,S_0)=w(t_0,{\cal Y}_{0})&=& \E^{\tilde{\bP}}(w(T,{\cal Y}_T))\nonumber\\
&=&\E^{\tilde{\bP}}\left( -\frac{1}{\cS_T({\cal Y}_T)}-\lambda\ds \Phi(\cS_T({\cal Y}_T)) \right)\label{first-rep-u},
\enq
where ${\cal Y}_{0}$ is such that $\dx w(0,{\cal Y}_{0})=\frac{1}{S_0}$,    
and then
\ben
u(0,S_0)&=& S_0\ds u(0,S_0) + \frac{1}{\lambda}\left({\cal Y}_{0} - \ln (S_0)\right).
\enn

If $u$ satisfies (\ref{contrainte-strict}), and $s\ds\sigma$ is bounded, one can  choose $\tilde{\bP}$ absolutely continuous with respect to $\bP$ such that $Y={\cal Y} \; \bP-a.s.$, and then
\beq
dW^{\tilde{\bP}}_t = dW^\bP_t   +\left(\frac{s\ds\sigma -\sigma}{1-\lambda\gamma} dt - s\ds \sigma  \right)dt.\label{driftdY}
\enq
In particular this holds true under the assumptions i) to v) of Theorem \ref{ubvbm}.
\end{theo}
We state right away an extension of this result in the constant volatility case, which will come as a corollary of Theorems \ref{local-rep-2} and \ref{local-rep}.
\begin{theo}[Constant volatility]\label{maintheo}
In addition to the assumptions of Theorem \ref{local-rep}, if $\sigma$ is constant and (\ref{finite0}) holds, then 
\begin{enumerate}

\item One can chose $\tilde{\bP}$ to be equal to $\bQ$ defined in Proposition \ref{Backlund Y},

\item One can compute $u(0,S_0)$ as follows:
\begin{itemize}
\item[Step 1] From $\Phi$ compute $\cS_T$ as stated in Proposition \ref{definverse}.
\item[Step 2] Find ${\cal Y}_0$ such that $$\frac{1}{S_0}=\E^{\tilde{\bP}}\left(\frac{1}{\cS_T({\cal Y_0}+\sigma W^{\tilde{\bP}}_T + \frac{\sigma^2}{2}T)}\right).$$
\item[Step 3]
Compute $$u(0,S_0)= S_0 \E^{\tilde{\bP}}\left(\ds{\Phi}(\cS_T({\cal Y_0}+\sigma W^{\tilde{\bP}}_T+ \frac{\sigma^2}{2}T))\right)  +\frac{1}{\lambda}({\cal Y_0}-\ln(S_0)).$$
\end{itemize}

Moreover there holds
\beq
u(0,S_0)=\E^\bP\left( S_T\ds{\Phi}(S_T) \right)  -\frac{1}{\lambda}(\ln(S_0)-{\cal Y_0})\label{modifBS}.
\enq

\item For all $\varphi$ such that,
$$\frac{\varphi(\cS_T)}{\cS_T}\exp\left(-\frac{y^2}{2\sigma^2T}\right)\in L^1,$$  there holds
\beq
&&\E^\bP \big(\varphi(S_T)\big) = S_0\E^{\tilde{\bP}}\Big(\frac{\varphi(\cS_T)}{\cS_T}(\sigma W^{\tilde{\bP}}_T+Y_0+ \frac{\sigma^2}{2}T)\Big)\label{ana1}\\
&=& S_0 \int_\R \varphi(s)\frac{1-\lambda s^2\partial_{ss}\Phi}{s^2}\exp(-\frac{(V_T(s)-\frac{\sigma^2}{2}T-Y_0)^2}{2\sigma^2T})\frac{ds}{\sigma \sqrt{2\pi T}}. \nonumber
\enq

\item If moreover  
\beq\label{finite2}
\int_{\R} \exp\left(-\frac{y^2}{2\sigma^2T}\right) \frac{\ln(\cS_T(y))}{\cS_T(y)} dy  \text{ is finite},
\enq then
\beq
\E^\bP\left(\int_0^T \frac{d\lb S,S \rb_t}{S^2_t}\right)&=&\E^\bP\left(\int_0^T\frac{\sigma^2}{(1-\lambda\gamma(t,S_t))^2}\right)\text{ is finite}\label{finitevariance},
\enq
and 
\beq
\E^\bP(\Phi(S_T))\text{ is finite}\label{finitephi}.
\enq

\end{enumerate}
\end{theo}

{\it Remarks.} The representation formula (\ref{first-rep-u})  is a Widder's type uniqueness result, as it shows that the solution of the  parabolic equation (\ref{w1}) (seen as a linear equation) is unique and defined by the representation formula (\ref{first-rep-u}). Again the key ingredient here is that we restrict ourselves to concave solutions (wich replaces the positivity assumption needed in Widder's Theorem.)

Regarding formulas (\ref{modifBS}) and why we call it a {\it modified Black-Scholes formula}, it is well known that the classical Black-Scholes formula for a call option consists of two terms 
\ben
u(0,S_0) &=& S_0{\cal N}(d_1) - K{\cal N}(d_0)\\
&=& S_0 \ds u(0,S_0) - K \bP\{S_T \geq K\}\\
&=& \E\left(S_T \ds \Phi(S_T)\right) - \E\left(S_T\ds \Phi(S_T) - \Phi(S_T)\right).
\enn
Indeed, for $\Phi=(S-K)^+$, $s\ds\Phi - \Phi=K\1_{S\geq K}$),
this identity becomes a tautology since in the Black-Scholes model, both $u(t,S_t)$ and $S_t\ds u(t,S_t)$ are martingales under the risk neutral probability. In the linear market impact model, there is a risk-neutral probability (the probability $\bP$ in our notations), and while $u$ is not the expectation of its final value (see formula \ref{dualverif}), hence not a martingale under $\bP$, when $\sigma$ is constant, $S_t\ds u(t,S_t)$ remains a martingale, and formula (\ref{modifBS}) reads similarly
\ben
u(0,S_0) = \E^{\bP}( S_T \ds \Phi(S_T) ) - (S_0\ds u(0,S_0) - u(0,S_0)).
\enn

Taking  $w(t,x)=\zeta (t,e^{x})$, $\zeta$ satisfies
\beq\label{BSxi}
\dt \zeta + \frac{\sigma^2(t,s)}{2} \xi^2 \partial_{\xi\xi} \zeta + \sigma^2\xi \partial_\xi \zeta=0,
\enq
with $s=( \xi \partial_\xi \zeta)^{-1}$. Then
\ben
\zeta(t, s e^{-\lambda u^c(t,s)}) = -\frac{1}{s} - \lambda\ds u(t,s),
\enn
with $u^c(t,s) = s\ds u - u$.
We now let $\Xi_t$ be the solution to 
\ben
\frac{d\Xi_t}{\Xi_t} &=& \sigma(t,{\cal S}_\lambda(t,\ln(\Xi_t)))dW^{\tilde{\bP}}_t + \sigma^2dt,\\
\Xi_{t_0} &=& S_0 e^{-\lambda u^c(t_0,S_0)},
\enn
and then $\zeta(t,\Xi_t)$ is a martingale, and a similar result hold. In the limit $\lambda \to 0$, we recover simply that $e^{\sigma W_t - \sigma^2/2t}$ is a martingale.

{\it Proof of Theorem \ref{local-rep}.} It has already been seen that $w(t,{\cal Y}_t)$ is a local martingale under $\tilde{\bP}$, and it remains to prove that is a martingale up to time $T$.
Arguing as in the proof of Lemma \ref{lemmedur}  (see section \ref{prooflemmedur} in Appendix) , we have
\ben
w(t,{\cal Y_t}) &\leq&  w(T,0)+ \dx w(T,0)({\cal Y_t}+\demi\bsig^2(T-t)),
\enn
therefore letting 
$$Z_t=w(t,{\cal Y_t})-\dx w(T,0) \int_0^t \sigma_r dW^{\tilde{\bP}}_r,$$ 
$Z_t$ can be bounded by above by a constant independent of $t$,
as $\sigma$ is bounded. Hence if $w(t,{\cal Y}_t)\wedge 0$ is equi-integrable, 
$(Z_t)_{t\geq 0}$ is equi-integrable, which  implies that $\E^{\tilde{\bP}}(Z_t) \equiv Z_0$.  
This is enough to conclude following standard arguments (see \cite{KaratzasShreve}) that $Z_t$ is a martingale, which in turn allows to conclude that $w(t,{\cal Y}_t)$ is a martingale up to time $T$.

The proof of 
\begin{lemme}\label{weqint}
Under assumption (\ref{kernel}), $w(t,{\cal Y}_t)\wedge 0$ is equi-integrable for $t\in [0,T]$.
\end{lemme}
is deferred to the appendix, section \ref{proofweqint}.

$\hfill \Box$

To prove the existence of ${\cal Y}_0$, from Proposition \ref{defilegendre}, under condition (\ref{kernel}) there holds
\begin{itemize}
\item[-] $\lim_{\xi \to -\infty} \dx w_T(x) = +\infty$,
\item[-] $\lim_{\xi \to +\infty}\dx w_T(x)=0$,
\end{itemize}
and $\dx w$ is non-decreasing,
which allows to conclude the existence and uniqueness of ${\cal Y}_0$.

$\hfill \Box$

{\it Proof of Theorem \ref{maintheo}.} This theorem is a corollary of Theorems \ref{local-rep-2} and \ref{local-rep}. For Point 1, once $S$ is a martingale up to $T$, one can define the probability $\bQ$, and  $Y$ follows then
\beq\label{Ysimple}
dY_t = \sigma dW^{\bQ}+\frac{\sigma^2}{2}dt, 
\enq
for $W^\bQ$ a $\bQ-$ Brownian motion.

For point 2, when $\sigma$ is constant, $\dx w$ follows also equation \ref{eqonw}, which shows the Step 2 of the Point 2, and one then uses formula (\ref{first-rep-u}).

Point 3 is the direct consequence of  (\ref{Ysimple}), 

Point 4 is a particular case of (\ref{ana1}) when $\varphi(s)=\ln(s)$. Note that as remarked at the end of the proof of Theorem \ref{local-rep-2}, the finiteness of $\E^\bP(\ln(S_T))$ is equivalent to the finiteness of $\E^{\bP}(\Phi(S_T))$.

$\hfill \Box$

\section{a Black-Scholes-Legendre formula}
In this section we derive the first order expansion of the solution with respect to the market impact parmeter $\lambda$.  To establish rigorously this expansion, we will need the regularity results of the pricing equation, and the representation formula previously established.

\subsection{Formal computations}
For this we will use the identity (\ref{verif}). We are in the case where $\lambda, \sigma$ are constant.
We still consider a probability $\bP$ under which $S$ follows (\ref{dSver1}, \ref{dSver2}, \ref{dSver3}), and we let $\bQ_{BS}$ be a probability under which $\frac{dS}{S}=\sigma dW^{\bQ_{BS}}$ for  $W^{\bQ_{BS}}$ a $\bQ_{BS}$ Brownian motion. Consider $\bar{u}$ the Black and Scholes solution, i.e. the solution to (\ref{main-lambda-dep1}) for $\lambda=0$,
a formal Taylor expansion around $\lambda=0$ of (\ref{main-lambda-dep1}) yields
\ben
v(0,S_0)=\bar{u}(0,S_0) + \frac{\lambda}{2} \E^{\bQ_{BS}}\left( \int_0^T\sigma^2 \bar\gamma^2 dt\right)+o(\lambda),
\enn
and $\bar \gamma = s^2\dss \bar{u}.$
We now evaluate ${\cal I} = \E^{\bQ_{BS}}\left( \int_0^T\sigma^2 \bar\gamma^2 dt\right)$. 
To do this, we define  
\beq\label{defLc}
\bar{u}^c(t,s)=s\ds \bar{u} - \bar{u},
\enq 
note that $$\bar{u}^c(t,s)=\bar{u}^*(t,\ds u(t,s)),$$
for $\bar{u}^*$ the Legendre transform of $\bar{u}$.
Observe that, if $\bar{u}$ solves the Black Scholes equation, then $\bar{u}^c$ is a martingale under $\bQ_{BS}$ (this is checked by a simple computation). Then $d \bar{u}^c=S_t\dss \bar{u} dS_t$. 

We have then, under ${\bQ_{BS}}$,
\ben
{\cal J}  &=& \int_0^T\bar{\gamma}\sigma dW^{\bQ_{BS}}_t\\
&=& \int_0^T S_t \dss \bar{u} dS_t\\
&=& \bar{u}^c(T,S_T)-\bar{u}^c(0,S_0).
\enn
Then
\ben
\E^{\bQ_{BS}}({\cal J}^2) &=& 
\E^{\bQ_{BS}}(\bar{u}^c(T,S_T)-\bar{u}^c(0,S_0))^2\\
&=&\E^{\bQ_{BS}}((\bar{u}^c(T,S_T)^2) - (\bar{u}^c(0,S_0))^2,
\enn
moreover, ${\cal I} = \E^{\bQ_{BS}}({\cal J}^2)$. Then we obtain a first modified Black-Scholes-{\it Legendre} formula
\beq\label{newBS}
u(0,S_0) &= &\E^{\bQ_{BS}}(\Phi) + \frac{\lambda}{2}\left(\E^{\bQ_{BS}}((\Phi^c)^2) - (\E^{\bQ_{BS}}(\Phi^c))^2\right)+o(\lambda),\\
\Phi^c &=& s\ds \Phi - \Phi.\nonumber
\enq
We then observe that, if $\Psi$ is any reasonable Lipschitz function and
\beq\label{defPhi}
\Phi=\inf\{\Psi'\geq \Psi, \lambda s^2\dss \Psi' \leq 1\}.
\enq
Then
\ben
\E^{\bQ_{BS}} (\Phi(S_T)) &=& \E^{\bQ_{BS}} (\Psi(S_T)) + O(\lambda^2),\\
\E^{\bQ_{BS}} (\Phi^c(S_T)) &=& \E^{\bQ_{BS}} (\Psi^c(S_T)) + O(\lambda).
\enn
Hence, replacing $\Phi$ by $\Psi$ in the above formula will only add terms of order 2 terms in $\lambda$. In the case of a call (resp. put) option with strike $K$ this yields $\Phi^c(s)=K\1_{s\geq K}$ (resp. $\Phi^c(s)=-K\1_{s\leq K}$). Now let  $$d_K=\E^{\bQ_{BS}}(\1_{s\geq K}),$$ 
$C_{BS}$ (resp. $P_{BS}$) be the Black Scholes call (resp. put) price,  then 
\beq\label{BSformula}
C_{BS}^\lambda(0,S_0) &=& C_{BS}(0,S_0) + \frac{\lambda}{2}K^2\left( d_K - d_K^2 \right)+o(\lambda),\\
P_{BS}^\lambda(0,S_0) &=& P_{BS}(0,S_0) + \frac{\lambda}{2}K^2\left( d_K - d_K^2 \right)+o(\lambda).\label{BSformulaput}
\enq
{\it Remarks.} Note that (\ref{newBS}) allows to compute  the first order market impact correction  using a simple Monte-Carlo/analytical pricer, for any terminal payoff as long as one is able to compute $\Phi^c$. It can be computed without relying on the constant volatility assumption, so it can still be computed in presence of local or even stochastic volatility, however, in those cases $u^c$ is not a martingale anymore, so the first order expansion should be slightly different.

Note also that the analytical price of call spread with strikes $K_1<K_2$ would be easily computed since in that case $\Phi^c=K_1\1_{K_1\leq s \leq K_2}- (K_2-K_1)\1_{s\geq K_2}$.

Note  that the  correction term is indeed quadratic in $\Phi$.

The correction obtained here is also formally valid for the case of pure liquidity costs studied by Cetin, Soner and Touzi \cite{CetinSonerTouzi}, since at first order the two pricing equations are the same, it also should be valid for any fully non-linear modification of the Black Scholes equation.

It now remains to turn the formal expansion into a rigorous statement.

\subsection{First order expansion of the solution}
\begin{theo}\label{theo-expa}
Let $\Psi$ be a terminal payoff and let $\Phi^\lambda$ be defined as in (\ref{defPhi}). Assume that $\Psi$ is globally Lipschitz and satisfies the assumptions i) to iv) of Theorem \ref{ubvbm}, and that $\Psi^c=s\ds \Psi - \Psi$ is bounded. Let $u(t,s)$ be the solution of (\ref{main-lambda-dep1}, \ref{main-lambda-dep2}, \ref{main-lambda-dep3}) with constant volatility parameter $\sigma$. 
Then $u$ is differentiable with respect to $\lambda$ and there holds
\beq\label{dlambda}
\frac{\partial u(0,S_0)}{\partial \lambda} = \E^\bP\left(\int_0^T \frac{\sigma^2}{2}\frac{\gamma^2}{(1-\lambda \gamma)^2}(t,S_t)dt\right), 
\enq
where $S_t$ follows (\ref{dSver1}, \ref{dSver2}, \ref{dSver3}), 
and moreover
\beq\label{newBS2}
u(0,S_0) &= &\E^{\bQ_{BS}}(\Psi) + \frac{\lambda}{2}\left(\E^{\bQ_{BS}}((\Psi^c)^2) - (\E^{\bQ_{BS}}(\Psi^c))^2\right)+o(\lambda).
\enq

\end{theo}
 
{\it Proof.}  
We first start with a smoothed terminal payoff $\Psi\ep $ satisfying  the constraint (\ref{contrainte-strict}).
Let $u_\lambda$ be the derivative of $u$ with respect to $\lambda$. Then
\ben
\dt u_\lambda + \frac{\sigma^2}{2}\frac{\gamma_\lambda}{(1-\lambda \gamma)^2} + \frac{\sigma^2}{2}\frac{\gamma^2}{(1-\lambda \gamma)^2}=0,
\enn
and $u_\lambda(T)=\partial_\lambda \Phi^\lambda$ is supported on $\{\lambda s^2\dss\Phi^\lambda=1\}$. 
One can thus write
\ben
u_\lambda(0,S_0) = \E^\bP\left(\int_0^T \frac{\sigma^2}{2}\frac{\gamma^2}{(1-\lambda \gamma)^2}(t,S_t)dt\right)+ \E^\bP(\partial_\lambda\Phi^\lambda(S_T)),
\enn
where $S_t$ is solution of (\ref{dSver1}).
We have that $\E^\bP(\partial_\lambda\Phi^\lambda(S_T))=0$ as it is supported on $\{\lambda s^2\dss\Phi=1\}$. Then in order to show (\ref{dlambda}), 
we need to show that this integral indeed converges to the expected limit, as $\Psi\ep$ converges to $\Psi$, in particular near $\lambda = 0$.  From Theorem \ref{ubvbm}, and given our assumptions on $\Psi$, there hold uniform bounds on $\gamma$ and $(1-\lambda\gamma)^{-1}$ on $[0,T']$ for $T' <T$. Hence the integral up to $T'<T$ will converge when $\Psi\ep$ goes to $\Psi$ thanks to the dominated convergence Theorem. We just need to show that 
\ben
\E^\bP\left(\int_{T'}^T \frac{\sigma^2}{2}\frac{\gamma^2}{(1-\lambda \gamma)^2}(t,S_t)dt\right)
\enn
converges to $0$ as $T'\to T$, uniformly with respect to $\lambda$ and to the terminal condition $\Psi$, as long as $\Psi$ satisfies the assumptions of the Theorem.  For this we will use the results of Theorem \ref{local-rep-2}.
\begin{lemme}\label{Girsanov}
Under the assumptions of Theorem \ref{theo-expa}, let $\bQ$ be defined as in Proposition \ref{Girsanovloc}. 
For any smooth $C^2$ function $\phi(t,y)$ such that  $S_t\phi(t,Y_t)$ is a  martingale under $\bP$ or $\phi(t,Y_t)$ is a martingale under $\bQ$  there holds
\ben
\dt \phi+\frac{\sigma^2}{2} (\partial_{yy}\phi+ \dy \phi)=0.\label{heatimath}
\enn
In particular this holds for
\ben
\imath(t,y) = -\frac{1}{\cS(t,y)},\\
\jmath(t,y)= \ds u(t,\cS(t,y)).
\enn
\end{lemme}

{\it Proof.} Under the assumptions of Theorem \ref{theo-expa}, the assumptions of Theorem \ref{local-rep-2} are satisfied, hence $S$ is a martingale up to time $T$, and one can define $\bQ$ as in Proposition \ref{Girsanovloc}.
Then of course $S_t S_t^{-1}$ is a martingale under $\bP$, hence $(\cS(t,Y_t))^{-1}$ is a martingale under $\bQ$, and as shown in Theorem \ref{local-rep}, $w = -\frac{1}{\cS} -\lambda \ds u(\cS)$ is also a martingale under $\bQ$, hence so is $\ds u(\cS)$.

$\hfill \Box$

We now observe that 
\ben
\frac{\dy \imath}{\imath} &=& -\frac{1}{1-\lambda\gamma},\\
\frac{\gamma^2}{1-\lambda\gamma^2}&=&\frac{1}{\lambda^2}(\frac{\dy \imath}{\imath}-1)^2,
\enn
and 
\beq\label{koiH}
 \dy(\imath + \lambda \jmath)&=&\imath.
\enq
We then compute
\ben
Q&=&\E^\bP\left(\int_{T'}^T \frac{\sigma^2}{2}\frac{\gamma^2}{(1-\lambda \gamma)^2}(t,S_t)dt\right)\\
&=& \E^\bP\left(\int_{T'}^T\frac{\sigma^2}{2}\frac{1}{\lambda^2}\left(\frac{\dy \imath}{\imath}-1\right)^2(t,Y_t)dt\right)\\
&=& \E^{\bQ}\left(\int_{T'}^T\frac{\sigma^2}{2}\frac{\imath^{-1}}{\lambda^2}\left( \dy \imath-\imath\right)^2(t,Y_0+\sigma W^{\bQ}_t + (\sigma^2/2)t)dt\right),
\enn
where in the third line we have used Theorem \ref{theo-main}, formula (\ref{ana1}). Then assuming that $\Psi^c$ is bounded, there exists $C$ such that
\beq\label{bdI}
e^{-y-\lambda C} \leq \imath(T,y) \leq e^{-y+\lambda C},
\enq
and this estimate is propagated by the heat equation, so that for another value of $C$ the bound (\ref{bdI}) holds on $[0,T]$. 
Plugging this into the above equality yields
\ben
Q &\leq & \E^\bQ\left( \int_{T'}^T\frac{C\sigma^2}{2\lambda^2}\left( e^y(\dy \imath-\imath)^2\right)(t,Y_0+\sigma W^\bQ_t + (\sigma^2/2)t)dt\right)\\
&\leq & \E^\bQ\left( \frac{C'\sigma^2}{2\lambda^2}\int_{T'}^T\left(\dy \imath-\imath\right)^2(t,Y_0+\sigma W^\bQ_t + (3\sigma^2/2)t)dt\right)
\enn
where we have used Girsanov's Theorem for the second line.
We now use (\ref{koiH}) which yields for $\jmath(t,y)=\ds u(t,\cS(t,y))$
\ben
Q&\leq & \E^\bQ\left( \frac{C'\sigma^2}{2}\int_{T'}^T\left(\dy\jmath\right)^2(t,Y_0+\sigma W^\bQ_t + (3\sigma^2/2)t)dt\right),
\enn
where $\jmath$ solves (by Proposition \ref{Girsanov})
\beq\label{hitjloc}
\dt \jmath + \frac{\sigma^2}{2}(\dyy \jmath + \dy \jmath)=0.
\enq
Then it follows that, letting 
\ben
\tilde{Y_t}&=& Y_0+\sigma W_t + (3\sigma^2/2)t,\\
H_{T',T} &=& \int_{T'}^T\dy\jmath(t',\tilde{Y_{t'}})\sigma dW_t,
\enn
we have, for another constant $C$, 
\ben
Q\leq  \E^\bQ\left( CH_{T',T}^2\right).
\enn
Moreover, we have using (\ref{hitjloc})
\ben
H_{T',T}=\jmath(T,\tilde{Y_T})-\jmath(T',\tilde{Y}_{T'}) - \int_{T'}^T\dy\jmath(t,\tilde{Y_t})\sigma^2dt,
\enn
so that by Cauchy Schwartz's inequality
\ben
\left| H_{T',T}-(\jmath(T,\tilde{Y}_T)-\jmath(T',\tilde{Y}_{T'}))\right|  \leq  \varepsilon \left(\int_{T'}^T(\dy\jmath)^2(t,\tilde{Y_t})dt\right)^{1/2}
 \enn
 for $\varepsilon = \sigma^2(T-T')^{1/2}$,
hence $Q \leq C'' (\E^\bP(J_{T',T}^2) + \varepsilon^2 Q)$, where $J_{T',T}=\jmath(T,\tilde{Y_T})-\jmath(T',Y_{T'})$. It remains to show that $\E^\bP\left((J_{T',T})^2\right)$ converges to 0 uniformly with respect to $\lambda$ as $T'\to T$. This is a straightforward consequence of the dominated convergence theorem, as $\jmath(t,y) = \ds u(t,\cS(t,y))$ is bounded under our assumptions.

$\hfill \Box$

\section{Numerical simulations}
In this section we present for illustration purposes a numerical implementation of the model. Another study of a numerical implementation scheme is provided in the companion paper \cite{BoLoZo2}. 
We propose the following numerical scheme:
We set $\varepsilon$ to a small constant ($\varepsilon = 10^{-3}$ in our applications), and divide the time interval $[0,T]$ into $N$ time intervals $[0,t_1,\hdots,t_N=T]$. We let $\Delta t = \frac{T}{N}$.
\begin{itemize}
\item[-] Define the truncated operator
\be
F\ep(\gamma,\bar\gamma) = \frac{\gamma}{\max\{1-\lambda \bar\gamma, \varepsilon\}} .
\en

\item[-] Initialize $i=N$.
\item[-] {\bf Terminal condition} Initialize $u(t_N) = \Phi(T)$

\item[-] {\bf  time loop} For  $i=N$ down to $i=1$
\item[-] initialize $v_1 = u(t_i)$.

\item[-]{\bf Non linear iterations} for $j$ in $[1 .. N_{it nl}]$ 

\begin{itemize}
\item[-]  solve  for $w$
\be
\frac{u(t_i) - w}{\Delta t} &=& - \demi\sigma^2  F(s^2\partial_{ss}w,s^2\partial_{ss}v_j),\\
\en

\item[-] Set $v_{j+1} = w $ if $j<N_{itnl}$ and do one more non-linear iteration or
\item Set $u(t_{i-1})=w$ and exit non-linear iterations loop

\end{itemize}

\item[-] iterate on the time step $i-1$ with $u(t_{i-1})$ set above.

\end{itemize}

Typically, the number of non linear iterations $N_{it nl}$ needed for convergence was small: $N_{it nl} = 3$ was enough in our numerical example.

For stability reasons, we use an implicit scheme in the non-linear iterations. An alternative to this method would be to enforce the constraint on the gamma at each time step, but we empirically observe that our method worked quite well.  As noticed above, with a constant volatility model, it is enough to enforce the upper bound on $\gamma$ on the terminal condition, but this fails to be true with a generic local volatility. 

We present some numerical simulations of the linear model in Figures \ref{tt}, \ref{ttt}, \ref{tttt}.

We present the cases of 100-strike put option Figure \ref{tt} , {\bf minus} a 100-strike put option Figure \ref{ttt} ,  and a 90-100 call spread Figure \ref{tttt}.
The numerical values used are $\lambda=5.0 \, 10^{-3}$, $\sigma=0.3$, no interest rates or dividends.
Note that in Figure \ref{tt} (resp. \ref{ttt}) the option sold is convex (resp. concave), while in Figure \ref{tttt}   the second derivative of the payoff changes sign. One sees that in all cases the market impact plays against the option's seller (the put sold is more expensive, and put bought is cheaper).

{\centering
\begin{table}[ht]
\begin{tabular}{cc}
\begin{subfigure}{0.4\textwidth}\centering\includegraphics[width=6cm]{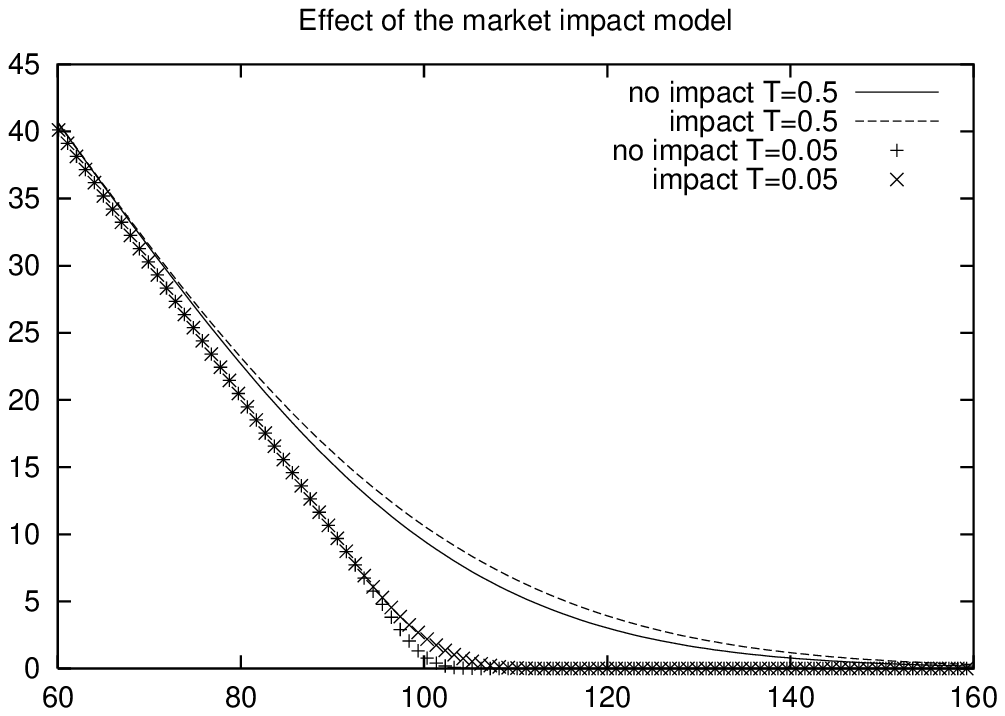}
\caption{The case a put option (gamma short case)}
\label{tt}\end{subfigure}&
\begin{subfigure}{0.4\textwidth}\centering\includegraphics[width=6cm]{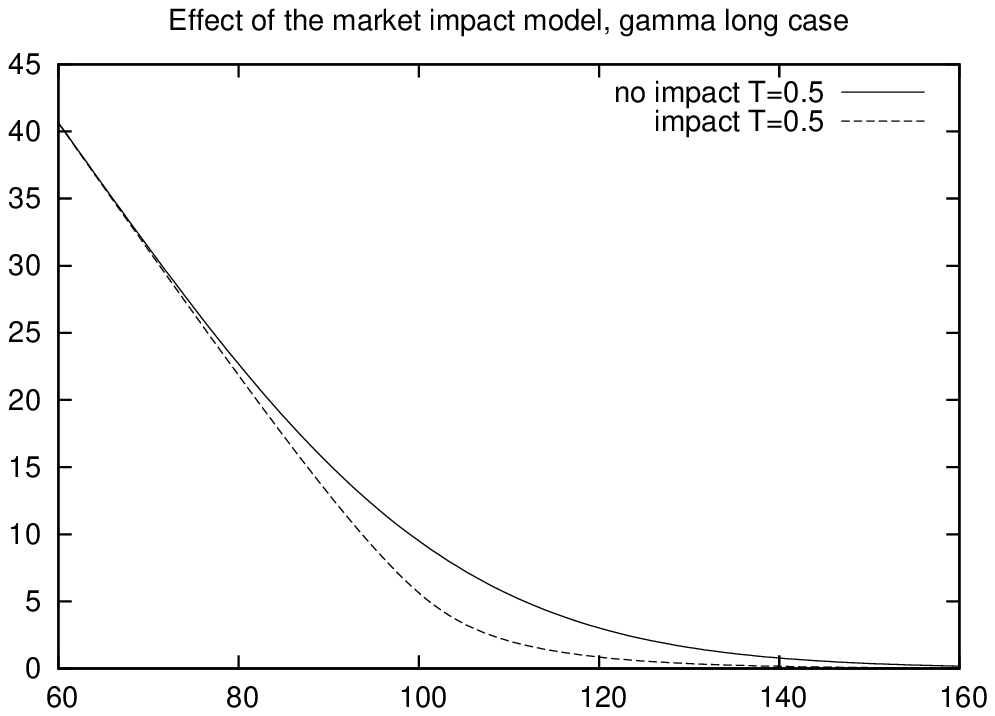}
\caption{The case of a put option(gamma long case)}
\label{ttt}\end{subfigure}\\

\begin{subfigure}{0.4\textwidth}\centering\includegraphics[width=6cm]{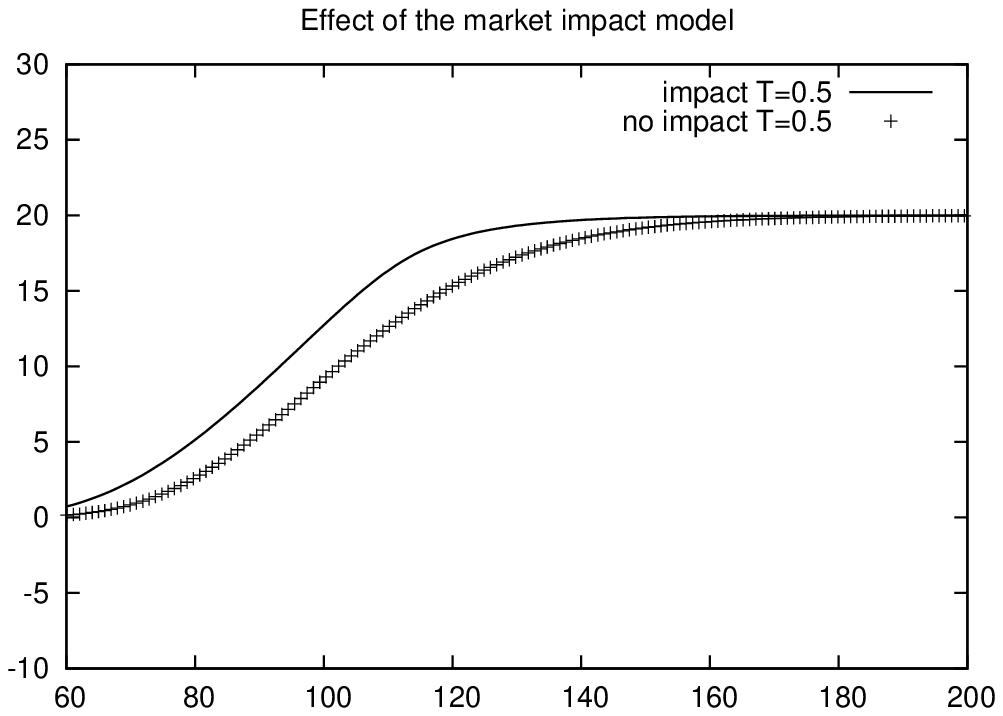}
\caption{The call spread case}
\label{tttt}\end{subfigure}&\\
\end{tabular}
\caption{Numerical Simulations}
\label{tab:mytable}
\end{table}
}

\newpage

\appendix

\section{}

\subsection{Proof of Theorem \ref{theo-comp}}\label{prooftheo-comp}
We perform the change of variable $s=e^y$, and consider $v_i(t,y)=u_i(t,e^y), i=1,2$, which yields that $s^2\dss u_i = \dyy v_i - \dy v_i$. Under the above assumptions,  $u_1$ and $u_2$ have logarithmic growth near $0$ and linear growth at $+\infty$, hence $v_1, v_2$ have  exponential growth. The derivatives of $v_i$ with respect to $y$ are bounded up to order 3, and $\sigma$ is uniformly Lipschitz as a function of $y$.
Then $w=v_1-v_2$ solves 
\ben
\dt w + \frac{\sigma^2(t,e^y)}{2}\frac{\dyy w - \dy w}{(1-\lambda(\dyy v_1-\dy v_1))(1-\lambda(\dyy v_2-\dy v_2))}=0.
\enn
This can be seen as a linear equation of the form
\ben
\dt w + a(t,y) \dyy w + b(t,y)\dy w=0,
\enn
where $a,b$ are bouded, uniformly Lipschitz in $y$, $a$ bounded away from 0. Then we can apply classical results of comparison under exponential growth (see \cite{Pascucci}).

$\hfill \Box$  

\section{}

\subsection{Proof of Lemma \ref{lemmedur}}\label{prooflemmedur}
We consider ${\cal Y}$ introduced in (\ref{defcalY}) under the probability $\tilde{\bP}$, with $W^{\tilde{\bP}}$ a $\tilde{\bP}$-Brownian motion, and $\E^{\tilde{\bP}}$ the expectation under $\tilde{\bP}$. We drop the superscripts $\tilde{\bP}$ for the rest of the proof proof.
We need first the following Lemma:
\begin{lemme}\label{lemme-tycho-cond}There exists $\kappa_0>0, C$ such that
\beq\label{tychocond} 
\forall \kappa \leq \kappa_0, \int_\R \exp\left({-\frac{x^2}{2C\kappa}}\right)\sup_{s\in [\kappa,T]}|w(t,x)|dx <+\infty.
\enq
\end{lemme}

{\it Proof of Lemma \ref{lemme-tycho-cond}.} First note that $w$ is concave, increasing and $\dx w$ has limit equal to 0 at $+\infty$. Then since $\sigma\leq \bsig$, 
\beq\label{decw}
\frac{d}{dt} \left(w(t, x+\demi \bsig^2 (t-T))\right) = -\demi \sigma^2 \dxx w +\demi(\bsig^2 -\sigma^2) \dx w \geq 0,
\enq
hence $t\to w(t,x+\demi \bsig^2 (t-T))$ is non decreasing. The concavity of $w$  implies then that $w(t,x+\demi \bsig^2 (t-T))-w_T(0) - x \dx w_T(0)$ is non-positive.
Then we consider ${\cal Y}_t$ defined in (\ref{defcalY}). Note that as we assume that $u$ is a classical solution, then $w\in C^{2,1}_{s,t,loc}$, and as moreover $\sigma$ is bounded,  ${\cal Y}$ is well defined. There holds that
\beq\label{defXt}
X_t=w(t,{\cal Y}_t) - w_T(0) - \dx w_T(0)({\cal Y}_t +\demi \bsig^2 (T-t))
\enq
is also non-positive and satisfies 
\ben
dX_t = (\dx w(t,{\cal Y}_t) - \dx w_T(0))\sigma_t dW_t + \dx w_T(0)\demi(\bsig^2-\sigma_t^2)dt.
\enn
It is therefore a non-positive sub-martingale, and there holds by Fatou's lemma
\ben
X_t &\leq & \E(X_{t+s}|{\cal F}_t) \\
& = &\E(w(t+s, {\cal Y}_{t+s}|{\cal F}_t)- w_T(0) - \dx w_T(0)\left(\E({\cal Y}_{t+s}|{\cal F}_t) +\demi \bsig^2 (T-t-s)\right).
\enn
This implies that $\E(w(t+s, {\cal Y}_{t+s}|{\cal F}_t)$ is finite. Then
 using the concavity and monotonicity of $w$ and It\^o's formula, one has
\beq\label{ineqforbw}
\E(w(t+s, {\cal Y}_{t+s}|{\cal F}_t) \leq \E(w(t+s, {\cal Y}_{t}+ \usig W_s + \demi \bsig^2 s)
\enq
which shows that for $t+s \leq T$,
\ben
\frac{1}{(2\pi s)^{\demi}}\int_\R \exp\left({-\frac{x^2}{2\usig^2 s}}\right)|w(t+s,x)|dx
\enn
is finite as long as $w(t,\cdot)$ is finite. The conclusion of the proof then follows by observing that, from (\ref{decw}), 
\ben
\sup_{s\in [t,T]}|w(s,x)| \leq  \sup_{s\in [t,T]}|w(t,x-\demi\sigma^2 (s-t))|.
\enn
 $\hfill \Box$

We now go back to the proof of Lemma \ref{lemmedur}.
Since $\dx w=\frac{1}{\cS}$, by standard manipulations on the heat kernel, (\ref{kernel}) implies that for some $\varepsilon>0$,
\beq\label{kernelbis}
\int_\R \exp\left(-\frac{x^2}{2\bar{\sigma}^2(T+\varepsilon)}\right) w_T(x)dx 
\enq
is finite, hence $\bar{w}$ is defined on $(-\varepsilon,T)$ and $\lim_{t\to T}w(t,x)=w_T$.
Observe that $\bar{w}$ (resp. $\underline{w})$ is solution to 
\ben
\dt \bw + \demi(\bar{\sigma}^2\dxx \bw + \underline{\sigma}^2\dx \bw)&=&0,\\
 \bw(T,\cdot)&=&w_T,
\enn
(resp.
\ben
\dt \uw + \demi(\usig^2\dxx \uw + \bsig^2\dx \uw)&=&0,\\
 \uw(T,\cdot)&=&w_T).
\enn

We write for $w$
\ben
\dt w + \demi \bsig^2 \dxx w + \demi \usig^2 \dx w = \demi(\bsig^2-\sigma^2) \dxx w - \demi(\sigma^2- \usig^2) \dx w.
\enn
We consider $S=\demi(\bsig^2-\sigma^2) \dxx w - \demi(\sigma^2- \usig^2) \dx w$, which is non-positive. 
From the concavity and monotonicity of $w$, one obtains by simple calculations that if $w$ satisfies (\ref{tychocond}) for some $\varepsilon_1$, then $\dx w$ and $\dxx w$ and hence $S$ also satisfy (\ref{tychocond}) for any $\varepsilon_2<\varepsilon_1$.
Consider then $\tilde{w}$ defined by
\ben
\tilde{w}(t,x) = -\int_{t}^T\frac{1}{\bsig\sqrt{2\pi(t'-t)}}\int_\R \exp\left(-\frac{(x+\demi \usig^2 (t'-t)-z)^2}{2\bsig^2(t'-t)}\right) S(t',z)dz dt'.
\enn
Then since $S(t',z)$ satisfies (\ref{tychocond}), $\tilde{w}$ is well defined for $t\in  [T-\tau, T]$ for some $\tau >0$, and solves
\ben
\dt \tilde{w} + \demi \bsig^2 \dxx \tilde{w} - \demi \usig^2 \dx \tilde{w}= S
\enn
with $\tilde{w}(T)=0$.
Finally we have 
\ben
w = \tilde{w} + \hat{w},
\enn
where $\hat{w}$ satisfies 
\beq\label{heatvep}
\dt \hat{w} + \demi \bsig^2 \dxx \hat{w} - \demi \usig^2 \dx \hat{w} = 0,
\enq
with $\hat{w}(T)=w_T$, and $\hat{w}$ satisfies (\ref{tychocond}).
Then we have the following lemma:
\begin{lemme}\label{lemmetycho}
There exists $\tau(\kappa)$ such that, on $[T-\tau, T]$,  there exists a unique solution of (\ref{heatvep}) that satisfies (\ref{tychocond}) with parameter $\kappa$.  
\end{lemme}
{\it Proof.} This is a straightforward adaptation of the proof of Tychonoff \cite{Tycho}. The inequality obtained there page 207 adapted to our case is
\ben
|w|(t,x) \leq  f(-R) \exp{-\frac{(x+R)^2}{2\bsig^2t}}\sqrt{\frac{t}{T}}+f(R) \exp{-\frac{(x-R)^2}{2\bsig^2t}}\sqrt{\frac{t}{T}},
\enn
with $f(A)=\sup_{0\leq t\leq T}\{|w(t,A)|\}$.
Choosing $t$ small enough so that the right hand side is integrable leads to the conclusion that $w(t,x)=0$.

$\hfill \Box$

Since $w$ and $\tilde{w}$ satisfy the growth condition (\ref{tychocond}),
Lemma \ref{lemmetycho} implies  that $\hat{w}$ is indeed equal to $\bar{w}$ defined above up to $T_\tau$.
The argument can then be repeated up to $T-n\tau$ and yields uniqueness up to $t=0$. Then $w = \tilde{w}+\bar{w}$, where $\tilde{w}$ is non negative,  hence we conclude $w \geq \bar{w}$. The other inequality $w \leq \uw$ follows directly from (\ref{ineqforbw}).

$\hfill \Box$

\subsection{Proof of Lemma \ref{weqint}}\label{proofweqint}
To prove the equi-integrability of $w(t,{\cal Y}_t)\wedge 0$, we will use Lemma \ref{lemmedur}, hence for all $K>0$ 
\ben
\E^{\tilde{\bP}}\left(w(t,{\cal Y_t})+K\right)^- &\geq & \E^{\tilde{\bP}}\left(\bar{w}(t,{\cal Y_t})+K\right)^- \\
&\geq & \E^{\tilde{\bP}}\left(\bw(t,{\cal Y}_0+ \bsig W_t + \frac{\usig^2}{2} t) +K\right)^-,
\enn
where the last line comes by using the concavity and the the monotonicity of $\bar{w}$.
As has been observed already, $\dx w(t,\cS) = \frac{1}{\cS}$, hence by standard computations on the heat kernel, the last line is uniformly bounded under (\ref{kernel}). Hence for $t\in [0,T]$, 
\ben
0 \geq \E^{\tilde{\bP}}\left(w(t,{\cal Y_t})+K\right)^-  \geq -\delta(K),
\enn where $\lim_{K\to +\infty} \delta(K) = 0$. By  Chebyshev's inequality, this implies that 
\ben
\E^{\tilde{\bP}}\left(K\1_{w\leq -2K} \right)   \leq  \delta(K),
\enn
hence that 
\ben
\E^{\tilde{\bP}}\left(K\1_{w\leq -K} \right)   \leq 2\delta(K/2),
\enn
Finally summing the two we have that
\ben
\E^{\tilde{\bP}}\left(w(t,{\cal Y_t})\1_{w\leq -K} \right)   \geq -(\delta(K) + 2\delta(K/2)).
\enn

$\hfill \Box$

\section{}
\subsection{Proof of Theorem \ref{theolocvol}}\label{prooftheolocvol}

Under the condition $\lambda s^2\dss \Phi \leq 1$,  $\lim_{s\to \infty} \ds \Phi={\cal L}\in \R \cup \{-\infty\}$. We first prove  the following Lemma:

\begin{lemme}\label{bound}
Let  $\Phi$ satisfy assumption (\ref{kernel}). 
For all $\tau>0$, for all $\kappa'>0$ small, there exists $(K, \kappa)(\tau,\kappa')$ , such that for $t'\in [0,T-\tau]$,
\ben
[\kappa', 1/\kappa'] \subset \dy v^*(t',[-K, K_{\cal L}])\subset [\kappa, 1/\kappa],
\enn
where 
\ben
K_{\cal L}=\left\{\begin{array}{ll}  {\cal L}-\frac{1}{K}\text{ if }{\cal L}\text{ is finite}\\K\text{ otherwise.}\end{array}\right.
\enn
 
The constants $K, \kappa$ depend only on the behaviour at $-\infty$ and $+\infty$ of $w_T$
\end{lemme}
{\it Proof of Lemma \ref{bound}}
We start by constructing two barriers as in the proof of Theorem \ref{ubvbm}: 
$\bar{v}, \underline{v}$ are the inverse of 
$\bar{w}, \underline{w}$  defined by (\ref{defbarw}, \ref{defuw}) in Lemma \ref{lemmedur}, and $\bw \leq w \leq \uw$, which then yields $\uv \leq v^* \leq \bv.$
From the construction of $w_T$ (see Proposition \ref{definverse}, point 8),
under assumption (\ref{kernel}) there holds
\ben
\lim_{s\to 0} \ln(s)+\lambda(\Phi - s\ds \Phi) &=& -\infty,\\
\lim_{s\to 0} -\frac{1}{s} - \lambda\ds \Phi &=& -\infty.
\enn
and moreover
\ben
w_T(\ln(s)+\lambda(\Phi - s\ds \Phi))&=& -\frac{1}{s} - \lambda\ds \Phi,\\
\dx w_T(\ln(s)+\lambda(\Phi - s\ds \Phi))&=& \frac{1}{s},
\enn
which imply that
\beq\label{lim1}
\lim_{x\to -\infty}  \dx w_T &=&+ \infty,\\
\lim_{x\to +\infty}  \dx w_T &=&0,\label{lim2}\\
\lim_{x\to+\infty} w_T &=& -\lambda {\cal L} \in \R \cup + \infty.\label{lim3}
\enq
Limits (\ref{lim1}, \ref{lim2}, \ref{lim3}) are propagated for time $t\leq T$ and, for $t<T$, $\bar w$ and $\underline{w}$ are strictly increasing.  
Moreover, as noted in (\ref{decw}), $t\to w(t,x+\frac{\bsig^2}{2}(t-T))$ is non decreasing, hence
\ben
\bw(0,x-\frac{\bsig^2}{2} t) \leq w(t,x) \leq \uw (T-\tau, x+\frac{\bsig^2}{2}(T-\tau-t)).
\enn
The concavity of $w$ and (\ref{lim1}, \ref{lim2}, \ref{lim3}) imply then a control on the way in which $w$ goes to $-\infty$ at $-\infty$ and goes to $+\infty$ as $+\infty$, which passing to $v^*$ yields the lemma.

$\hfill \Box$

{\it Proof of Theorem \ref{theolocvol}.} We  study (\ref{eqonv*}) on compact sets of $(\infty, -\lambda{\cal L}) \times [0,T-\tau]$ for small $\tau$.  
From Lemma \ref{bound}, we see that (\ref{eqonv*}) is uniformly parabolic and that we have an a priori bound for $v^*$ and $\dy v^*$.  We can thus apply Lemma \ref{Lieberman}, to obtain local $C^{1+\alpha}_x$ regularity for $v^*$ for time $t<T$.
As in the proof ot Theorem \ref{ubvbm}, equation (\ref{eqonv*}) on $v^*$ can now be seen as a linear parabolic equation with H\"older coefficients, and Schauder estimates then yield that $v^* \in C^{1+\alpha/2, 2+\alpha}_{t,x,loc}([0,T)\times (-\infty, -\lambda{\cal L}))$. By differentiating the equation, further regularity follows if $\sigma$ has additional regularity. Then using the left side of the inequality in Lemma \ref{bound}, the local regularity of $v^*$ on $(-\infty, -\lambda{\cal L} )$ implies local regularity of $v$ on $\R_+^*$, henceforth of $u$ on any compact set of $[0,T)\times \R_+^*$. This achieves the proof of Theorem \ref{theolocvol}.

$\hfill \Box$

\section{Existence of smooth solutions for general $\lambda$}\label{genlambda}
Here we state a simple result of smoothness for "good" initial data in the case where $\lambda = \lambda(\gamma)$, when $\sigma$ is constant. In this case, we need to start with well behaved solutions, as the singularity of the solution can not be treated by the Legendre transform technique of the previous section. What we show is the following:
\begin{theo}
Let the final payoff $\Phi$ satisfy 
\beq\label{contrainte-lambda-dep}
-\frac{1}{\varepsilon} \leq\lambda s^2 \dss \Phi \leq 1-\varepsilon
\enq
 for some $\varepsilon>0$. 
Assume that $\lambda(\gamma)$  is smooth satisfies locally uniformly with respect to $\gamma$ the condition (\ref{ellip-lambda}). 
Then there exists a smooth solution to (\ref{main-lambda-dep1}, \ref{main-lambda-dep2}, \ref{main-lambda-dep3}) such that (\ref{contrainte-lambda-dep}) is satisfied for all time, that belongs to $C^{2+\alpha, 1+\alpha/2}_{s,t}([0,T)\times \R_+^*)$.
\end{theo}

{\it Proof.} We only sketch the proof, this is an adaptation of Propositon \ref{exiphi} that relies on the estimate formula \ref{FKV}.
This is a simple consequence of the identity (\ref{FKV}). From this and the conditions on the boundary data $\Phi$, one can construct a solution such that (\ref{FKV}) holds and $\gamma$ remains  bounded above and below. Then classical parabolic regularity theory yields the result.

$\hfill \Box$

We make the following observation:
\begin{prop}
Let $0\in I$, $F:\gamma\in I\to \R$ be a smooth increasing function such that $F(0)=0$ and $F(\gamma)\geq \gamma$. Then there exists $\lambda(\gamma)\geq 0$ such that
\ben
F(\gamma) = \frac{\gamma}{1-\lambda(\gamma)\gamma}.
\enn
Moreover $\lambda=\frac{1}{\gamma}-\frac{1}{F}$.
\end{prop}
This comes by elementary computations, and shows that under mild conditions on $F$, any fully non-linear pde of the form $\dt u + \demi\sigma^2F(s^2\dss u)=0$ can be derived as a market impact pricing equation. Note that the conditions imply 
$F'(0)=1$ and $F''(0)\geq 0$.

\bibliography{finance}

\vspace*{2cm}
\noindent Gregoire Loeper\\
Monash University\\
School of Mathematics\\
9 Rainforest Walk\\
3800 CLAYTON VIC, AUSTRALIA\\
\\
email: gregoire.loeper@monash.edu

\end{document}